\documentclass[aps,prb,reprint,twocolumn,superscriptaddress]{revtex4-2}

\usepackage{graphicx}
\usepackage{dcolumn}
\usepackage{bm}
\usepackage{latexsym,epsfig}
\usepackage{graphicx}
\usepackage{verbatim}
\usepackage{comment}
\usepackage{amsmath}
\usepackage{amssymb}
\usepackage{physics}
\usepackage{stmaryrd}
\usepackage{color}
\usepackage{epstopdf}
\usepackage{grffile}
\usepackage{lipsum}
\usepackage{enumitem}
\usepackage{ulem}
\usepackage[dvipsnames]{xcolor}
\DeclareGraphicsExtensions{.eps}

\newcommand{\beq}{\begin{equation}}
\newcommand{\eeq}{\end{equation}}
\newcommand{\bea}{\begin{eqnarray}}
\newcommand{\eea}{\end{eqnarray}}
\newcommand{\ben}{\begin{eqnarray*}}
\newcommand{\een}{\end{eqnarray*}}
\newcommand{\bfig}{\begin{figure}}
\newcommand{\efig}{\end{figure}}

\usepackage{hyperref}
\hypersetup{
    colorlinks=true,      
    urlcolor=blue,
    citecolor=blue,
    linkcolor=blue
}

\usepackage{booktabs}
\usepackage{array}
\usepackage{lipsum} 
\usepackage{array} 

\begin{document}
\title{Interplay of Quasiperiodicity, Hubbard Interactions, and Staggered Magnetism in a 1D Ring: Localization and Re-entrant Characteristics}
\author{Souvik Roy}
\email{souvikroy138@gmail.com}
\affiliation{School of Basic Sciences, Indian Institute of Technology Bhubaneswar, Odisha 752050, India}
\author{Ranjini Bhattacharya}
\email{ranjinibhattacharya@gmail.com}
\affiliation{Institute of Physics, Sachivalaya Marg, Bhubaneswar-751005, India}
\affiliation{Homi Bhabha National Institute, Training School Complex, Anushaktinagar, Mumbai 400094, India}

\date{\today}

\begin{abstract}
We study interaction-induced localization in a spinful quasiperiodic Hubbard ring with quasiperiodically modulated hopping and a staggered spin-dependent Zeeman field using a self-consistent Hartree-Fock approach. We characterize localization through the inverse participation ratio, normalized participation ratio, and multifractal analysis, and further examine its signatures in self-consistent real-space observables, including double occupancy, density fluctuations, local entropy, spin-density-wave order, and related correlation measures. As the Hubbard interaction is increased, we find a nonmonotonic evolution of the eigenstates: an initially extended regime gives way to an intermediate localized phase, followed by reentrant delocalization at stronger interactions. The width of this intermediate regime grows systematically with both the quasiperiodic hopping amplitude and the staggered Zeeman-field strength, indicating a cooperative enhancement of localization by these two ingredients. In the localized regime, the self-consistent interaction produces pronounced spin dependence in both the spectra and localization properties, resulting in distinct responses of the spin-up and spin-down sectors. The interaction-driven crossover is consistently captured by the real-space observables and further corroborated by real-time wave-packet dynamics, which reveal successive ballistic, confined, and reentrant-transport regimes consistent with the underlying eigenstate character. These results provide a unified framework for understanding the interplay of quasiperiodicity, electronic interactions, and spin-dependent fields in controlling localization, criticality, and quantum transport in correlated quasiperiodic systems.

\end{abstract}


\maketitle

\section{Introduction}


Understanding the interplay between disorder, quasiperiodicity, and electronic interactions remains a central problem in condensed-matter physics. Spatial inhomogeneity can profoundly modify quantum transport by controlling the spatial extent of single-particle wave functions. In noninteracting systems, random disorder can induce Anderson localization~\cite{r1,r2,r3}, whereby quantum interference suppresses coherent propagation and produces spatially localized eigenstates. Localization can also arise in quasiperiodic systems~\cite{r4,r5,r6,r7}, where deterministic, nonperiodic modulations generate localization without stochastic disorder. In contrast to conventional Anderson localization, quasiperiodic systems can exhibit sharp localization transitions controlled by the modulation strength, providing a distinct route from extended to localized quantum states.

Beyond the conventional extended and localized phases, quasiperiodic lattices~\cite{r8,r9,r10,r11,r12,r13,r14,r15,r16,r17,r18,r19,r20} can host critical eigenstates characterized by multifractal spatial structure. Such states form an intermediate regime between extended and exponentially localized behavior, with anomalous scaling properties that distinguish them from both limiting cases. The interplay among extended, critical, and localized states gives rise to unconventional spectral structures and nontrivial transport and dynamical responses. This broad phenomenology has established quasiperiodic systems~\cite{r21,r22,r23,r24,r25,r26,r27,r28,r29,r30,r31,r32,r33} as a versatile platform for studying localization transitions, mobility edges, and quantum critical behavior beyond the conventional framework of random disorder.

The presence of electron-electron interactions~\cite{r34,r35,r36,r37,r38} substantially enriches the localization problem, while their role in quasiperiodic systems remains incompletely understood. In contrast to noninteracting settings, where localization is governed primarily by single-particle interference, interactions compete with the quasiperiodic structure and modify the underlying many-body state~\cite{rsvk2,r39,r40,r41,r42}. Depending on the interaction strength and modulation parameters, correlations can enhance localization or instead promote delocalization by redistributing spectral weight and renormalizing the effective local potential. Such competition can produce nontrivial localization-delocalization crossovers, reentrant phases, and extended critical regimes. Establishing the microscopic mechanisms underlying these behaviors, together with their manifestations in spectral, transport, and dynamical observables, is therefore essential for a comprehensive understanding of correlated quasiperiodic systems.

The localization problem becomes substantially richer when spin degrees of freedom are incorporated. Spin-dependent fields, such as Zeeman splittings and spatially modulated magnetic textures~\cite{rsvk}, break spin degeneracy and reshape the single-particle spectrum, allowing the two spin sectors to exhibit distinct localization and transport responses~\cite{r43,r44,r45,r46}. In the presence of electronic correlations, spin polarization further feeds back into the effective potential, generating self-consistent spin-dependent fields that can strongly modify the spatial structure of the eigenstates. As a result, even a modest spin imbalance can be amplified by interactions, producing pronounced asymmetries between opposite-spin sectors in their spectral and localization properties. These effects provide a route to spin-selective control of quantum transport in correlated quasiperiodic systems and are relevant to spintronic and spin-caloritronic architectures, where controllable spin-dependent responses are essential.

Motivated by these considerations, we study a spinful Hubbard ring with quasiperiodically modulated hopping subject to a staggered spin-dependent Zeeman field. Electronic correlations are treated within a self-consistent Hartree-Fock scheme~\cite{r47,r48,r49,r50,r51}, which incorporates the interaction-induced renormalization of the effective single-particle Hamiltonian while retaining computational tractability for relatively large systems. Although mean-field theory does not capture quantum fluctuations beyond the self-consistent level, it provides a controlled framework for examining the coupled effects of quasiperiodicity, interactions, and spin-dependent fields. Within this approach, we systematically investigate how correlations reshape the energy spectrum, localization properties, and spin-resolved electronic structure, providing a unified basis for analyzing both the equilibrium and nonequilibrium dynamics of the correlated quasiperiodic system.

To systematically disentangle the roles of quasiperiodicity, electronic correlations, and spin-dependent fields, we employ a set of complementary diagnostics that goes beyond conventional spectral characterization. Localization is quantified using the inverse participation ratio and normalized participation ratio, supplemented by combined localization measures that differentiate extended, critical, and localized states. The scaling structure of the eigenstates is examined further through multifractal analysis, using the averaged fractal dimension as a quantitative measure of criticality. We also investigate self-consistent real-space observables-including charge-density fluctuations, local spin polarization, double occupancy, excitation gaps, and site-resolved entropy, to elucidate how interactions modify the underlying electronic structure. These equilibrium probes are complemented by real-time wave-packet dynamics, allowing the static localization properties of the eigenstates to be directly connected with their dynamical transport response. Together, the spectral, real-space, and dynamical analyses provide a comprehensive framework for identifying the mechanisms underlying interaction-driven localization and spin-selective transport in correlated quasiperiodic systems.

Our results uncover a nonmonotonic interaction-driven evolution of the electronic phases. The system is extended at weak interaction, but increasing the Hubbard coupling initially promotes localization. Upon further increasing the interaction strength beyond a characteristic scale, this trend reverses and the system reenters a delocalized regime, demonstrating that correlations can alternatively hinder or restore coherent transport. The intermediate critical regime also expands systematically with the quasiperiodic hopping amplitude and the staggered Zeeman-field strength, indicating that these two ingredients jointly stabilize critical behavior over an extended interaction range. In this regime, the two spin sectors acquire distinct spectral and localization characteristics, manifested by spin-dependent spectral splitting and contrasting inverse participation ratios. The associated reconstruction of the electronic states is further reflected in the self-consistent real-space observables, including charge and spin profiles, double occupancy, excitation gaps, and local entropy. These results establish a direct connection between the interaction-induced restructuring of the spectrum and the resulting real-space electronic response.

The paper is organized as follows. Section~II introduces the model and the self-consistent mean-field formulation. Section~III presents the main results, focusing on spectral reconstruction, localization measures, phase diagrams, and real-time wave-packet dynamics. Section~IV concludes with a summary of the principal findings and a discussion of their broader implications for interacting quasiperiodic systems and experimentally accessible platforms.

\section{Theoretical Formulation}
\begin{figure}[t]
    \centering
   \includegraphics[width=1.0\linewidth]{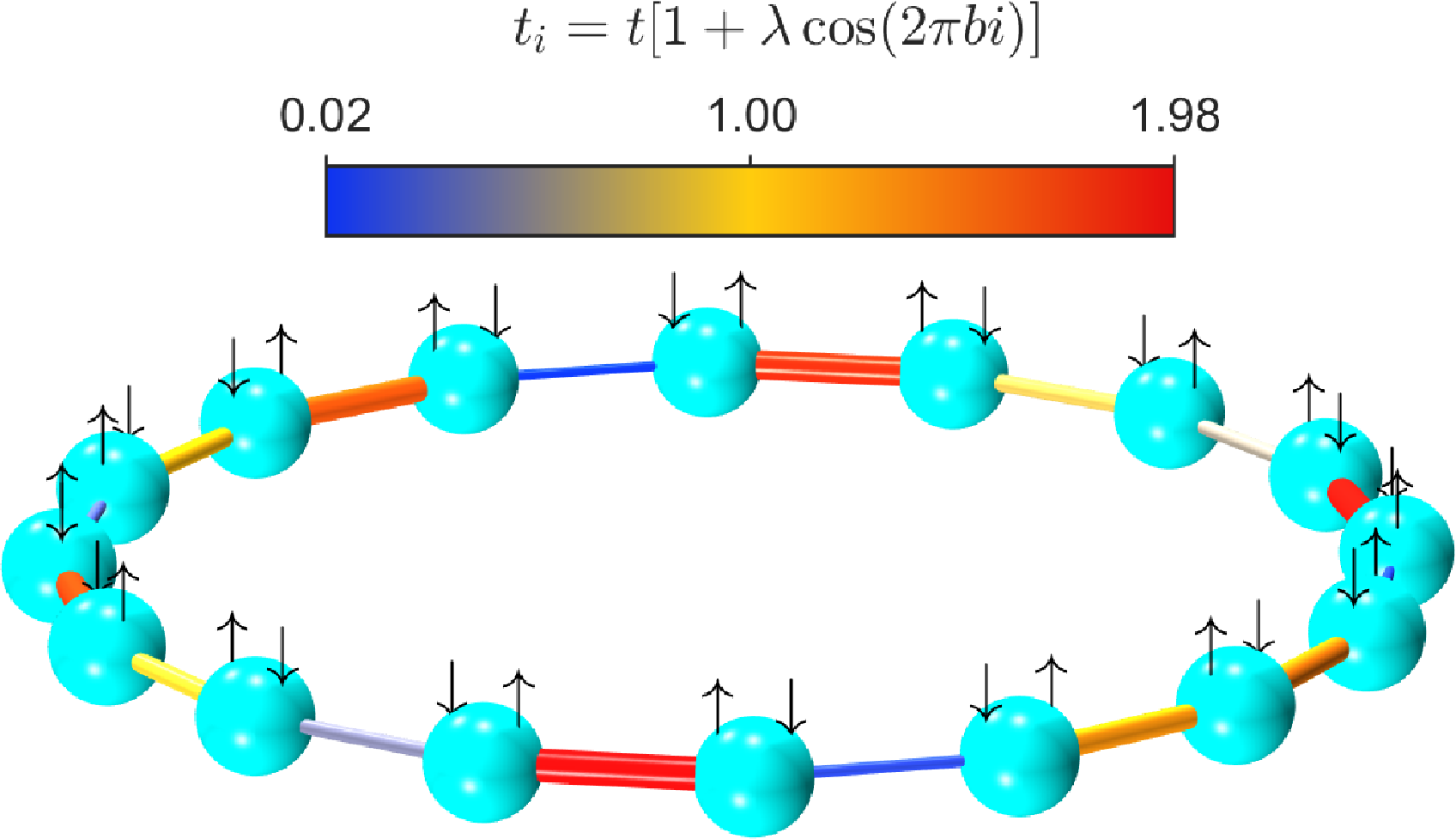}
    \caption{Schematic illustration of a spinful antiferromagnetic lattice with incommensurate hopping amplitudes, represented using distinct colors.}
    \label{fig:schm}
\end{figure}
In this section, we formulate the model Hamiltonian and describe the self-consistent mean-field approach adopted throughout the analysis. We then define the complementary diagnostics used to characterize localization, encompassing spectral properties, eigenstate-based measures, and real-space observables. Taken together, these quantities provide a systematic characterization of the localization behavior and its evolution with the relevant model parameters.
\subsection{Model and Mean-Field Formulation}

We consider a one-dimensional spinful Hubbard ring comprising $L$ sites, with quasiperiodically modulated nearest-neighbor hopping and a staggered Zeeman field that couples oppositely to the two spin species. The Hamiltonian is written as
\begin{equation}
\mathcal{H}
=
\mathcal{H}_{\mathrm{kin}}
+
\mathcal{H}_{U}
+
\mathcal{H}_{Z},
\end{equation}
with periodic boundary conditions.

The kinetic contribution is
\begin{equation}
\mathcal{H}_{\mathrm{kin}}
=
\sum_{i,\sigma}
t\left[1+\lambda\cos(2\pi b i)\right]
\left(
c_{i+1,\sigma}^{\dagger}c_{i,\sigma}
+\mathrm{H.c.}
\right),
\end{equation}
where $c_{i,\sigma}^{\dagger}$ ($c_{i,\sigma}$) creates (annihilates) an electron of spin $\sigma=\uparrow,\downarrow$ at site $i$. Here, $t$ is the reference hopping amplitude, $\lambda$ specifies the modulation strength, and $b$ is an irrational number that renders the hopping sequence quasiperiodic.

The on-site electron-electron interaction is described by the Hubbard term
\begin{equation}
\mathcal{H}_{U}
=
U\sum_i n_{i,\uparrow}n_{i,\downarrow},
\end{equation}
where $U$ denotes the interaction strength and
$n_{i,\sigma}=c_{i,\sigma}^{\dagger}c_{i,\sigma}$ is the local occupation operator. Spin degeneracy is explicitly lifted by a staggered Zeeman field,
\begin{equation}
\mathcal{H}_{Z}
=
h_z\sum_i(-1)^i
\left(
n_{i,\uparrow}-n_{i,\downarrow}
\right),
\end{equation}
where $h_z$ is the field amplitude. Thus, neighboring sites experience opposite Zeeman shifts, with the two spin species acquiring opposite on-site potentials. The resulting spin polarization is imposed externally through the staggered field rather than generated by spontaneous symmetry breaking.

We treat the interaction within a self-consistent Hartree-Fock approximation restricted to the density channel, consistent with a collinear spin configuration. The on-site interaction is decoupled as~\cite{r47,r48,r49,r50,r51}
\begin{equation}
n_{i,\uparrow}n_{i,\downarrow}
\approx
\langle n_{i,\uparrow}\rangle n_{i,\downarrow}
+
\langle n_{i,\downarrow}\rangle n_{i,\uparrow}
-
\langle n_{i,\uparrow}\rangle
\langle n_{i,\downarrow}\rangle .
\end{equation}
Here, the local densities $\langle n_{i,\sigma}\rangle$ are determined self-consistently from the occupied single-particle eigenstates. The final term contributes only a scalar energy shift and therefore does not enter the effective single-particle eigenvalue problem.

The resulting mean-field Hamiltonian can be expressed as
\begin{equation}
\mathcal{H}_{\mathrm{MF}}
=
\mathcal{H}_{\uparrow}^{\mathrm{eff}}
+
\mathcal{H}_{\downarrow}^{\mathrm{eff}}
+
E_0,
\end{equation}
where
\begin{equation}
E_0
=
-U\sum_i
\langle n_{i,\uparrow}\rangle
\langle n_{i,\downarrow}\rangle .
\end{equation}

The effective Hamiltonians for the two spin sectors are
\begin{equation}
\begin{aligned}
\mathcal{H}_{\uparrow}^{\mathrm{eff}}
=&
\sum_i
t\left[1+\lambda\cos(2\pi b i)\right]
\left(
c_{i+1,\uparrow}^{\dagger}c_{i,\uparrow}
+\mathrm{H.c.}
\right)
\\
&+
\sum_i
\left[
U\langle n_{i,\downarrow}\rangle
+h_z(-1)^i
\right]n_{i,\uparrow},
\end{aligned}
\end{equation}
and
\begin{equation}
\begin{aligned}
\mathcal{H}_{\downarrow}^{\mathrm{eff}}
=&
\sum_i
t\left[1+\lambda\cos(2\pi b i)\right]
\left(
c_{i+1,\downarrow}^{\dagger}c_{i,\downarrow}
+\mathrm{H.c.}
\right)
\\
&+
\sum_i
\left[
U\langle n_{i,\uparrow}\rangle
-h_z(-1)^i
\right]n_{i,\downarrow}.
\end{aligned}
\end{equation}

Although the two effective spin Hamiltonians are formally decoupled for a fixed self-consistent configuration, they remain coupled through their local densities. Specifically, the density of one spin sector determines the interaction-induced Hartree potential experienced by the opposite sector. Iterative solution of these coupled self-consistency conditions therefore generates site- and spin-dependent effective potentials. The resulting single-particle problem incorporates the quasiperiodic hopping modulation, interaction-induced Hartree fields, and staggered Zeeman potential on an equal footing. This framework consequently enables a systematic investigation of spin-resolved spectral reconstruction, localization properties, and nonequilibrium dynamics.

\subsection{Localization and Fractal Measures}

Once the Hartree-Fock equations have converged, the interacting problem is mapped onto two self-consistent effective single-particle Hamiltonians,
$\mathcal{H}_{\uparrow}^{\mathrm{eff}}$ and
$\mathcal{H}_{\downarrow}^{\mathrm{eff}}$, for the two spin sectors. The corresponding eigenvalue problem is
\begin{equation}
\mathcal{H}_{\sigma}^{\mathrm{eff}}
\ket{\psi_{n}^{(\sigma)}}
=
E_{n}^{(\sigma)}
\ket{\psi_{n}^{(\sigma)}},
\qquad
\sigma=\uparrow,\downarrow,
\end{equation}
where $E_{n}^{(\sigma)}$ denotes the $n$th eigenenergy and
$\ket{\psi_{n}^{(\sigma)}}$ is the associated normalized eigenstate. In the site basis, the eigenstate is represented by the amplitudes
$\psi_{n,i}^{(\sigma)}$ on sites $i=1,\ldots,L$, satisfying
\begin{equation}
\sum_{i=1}^{L}
\left|\psi_{n,i}^{(\sigma)}\right|^2=1.
\end{equation}
These self-consistent eigenstates provide the starting point for quantifying the spatial character of the spectrum through localization and multifractal measures.

\subsubsection{Inverse Participation Ratio and Normalized Participation Ratio}

The spatial character of the self-consistent eigenstates is first quantified using the inverse participation ratio (IPR)~\cite{IPR},
\begin{equation}
\mathrm{IPR}_{n}^{(\sigma)}
=
\sum_{i=1}^{L}
\left|
\psi_{n,i}^{(\sigma)}
\right|^{4}.
\end{equation}
For an extended state, the wave-function weight is distributed over the entire system and $\mathrm{IPR}_{n}^{(\sigma)}$ scales as $L^{-1}$. In contrast, for a spatially localized state, the IPR remains finite as $L\rightarrow\infty$. Thus, the IPR provides a direct measure of the spatial confinement of an eigenstate.

As a complementary measure, we consider the normalized participation ratio (NPR)~\cite{NPR}, defined by
\begin{equation}
\mathrm{NPR}_{n}^{(\sigma)}
=
\frac{1}
{L\,\mathrm{IPR}_{n}^{(\sigma)}}.
\end{equation}
The NPR approaches a finite value of order unity for an extended state, whereas it tends to zero for a localized state in the thermodynamic limit. The combined behavior of the IPR and NPR therefore enables a distinction between extended, localized, and intermediate states.

\subsubsection{Spectral Averaging}

To characterize the localization properties of an entire spin-resolved spectrum, we perform an average over all $L$ eigenstates,
\begin{equation}
\left\langle \mathrm{IPR} \right\rangle_{\sigma}
=
\frac{1}{L}
\sum_{n=1}^{L}
\mathrm{IPR}_{n}^{(\sigma)},
\qquad
\left\langle \mathrm{NPR} \right\rangle_{\sigma}
=
\frac{1}{L}
\sum_{n=1}^{L}
\mathrm{NPR}_{n}^{(\sigma)}.
\end{equation}
Here, $\sigma=\uparrow,\downarrow$ labels the spin sector. These spectral averages provide global measures of the localization and delocalization tendencies of the corresponding self-consistent spectra.

\subsubsection{Combined Localization Indicator}

To capture the simultaneous presence of localized and extended characteristics within the spectrum, we introduce the combined indicator
\begin{equation}
\eta_{\sigma}
=
\log_{10}
\left[
\left\langle \mathrm{IPR} \right\rangle_{\sigma}
\left\langle \mathrm{NPR} \right\rangle_{\sigma}
\right].
\end{equation}
Since the IPR decreases with increasing spatial extension whereas the NPR exhibits the opposite trend, their product provides a useful measure of the coexistence of the two characteristics. Consequently, $\eta_{\sigma}$ remains suppressed when the spectrum is predominantly localized or extended and becomes comparatively enhanced when localized and extended components coexist, as expected in an intermediate or critical regime.

\subsubsection{Fractal Dimension}

The spatial scaling of the eigenstates is further characterized through the second-order generalized fractal dimension~\cite{D2},
\begin{equation}
D_{2,n}^{(\sigma)}
=
-\frac{
\log\left[\mathrm{IPR}_{n}^{(\sigma)}\right]
}{
\log L
}.
\end{equation}
For a normalized state in a one-dimensional system, an extended state gives $D_{2}\simeq1$, whereas a localized state yields $D_{2}\simeq0$. Intermediate values of $D_2$ signal nontrivial spatial scaling and are characteristic of critical or multifractal states.

We characterize the global fractal properties of each spin sector through the spectral average
\begin{equation}
\left\langle D_2 \right\rangle_{\sigma}
=
\frac{1}{L}
\sum_{n=1}^{L}
D_{2,n}^{(\sigma)}.
\end{equation}
This quantity provides a compact measure of the overall spatial complexity of the spin-resolved eigenstates and allows the evolution of criticality to be tracked across the relevant parameter space.

All localization and fractal measures are evaluated using the fully converged eigenstates of the self-consistent effective Hamiltonians. Consequently, the effects of electronic interactions are incorporated directly through the self-consistent spin- and site-dependent Hartree potentials.

The remaining self-consistent real-space observables and the quantities used to characterize the real-time wave-packet dynamics are defined in the following subsections.
\begin{figure}[t]
    \centering
   \includegraphics[width=1.0\linewidth]{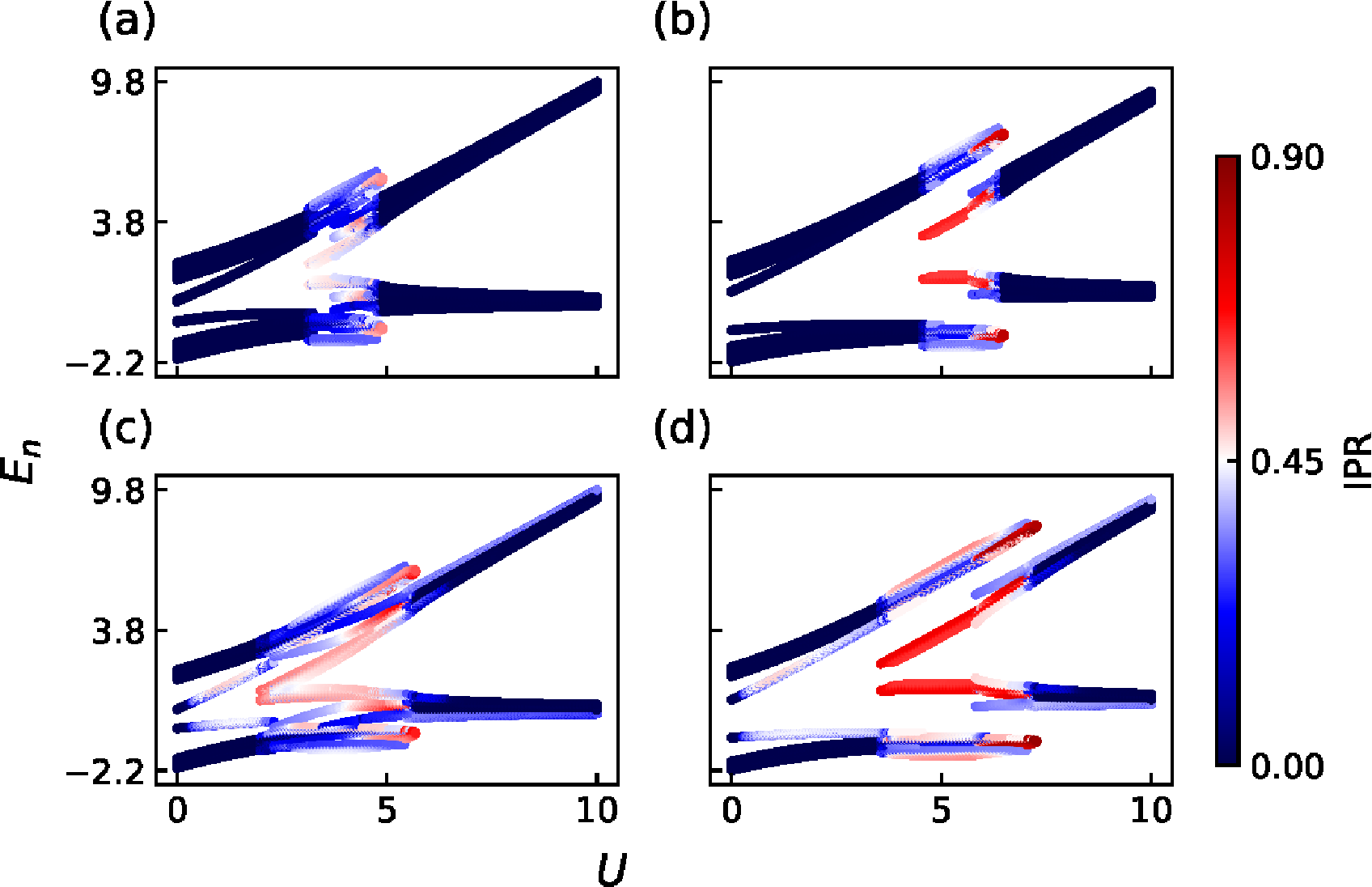}
    \caption{Combined spin-resolved mean-field eigenvalue spectrum as a function of the Hubbard interaction $U$.
For each value of $U$, all spin-up and spin-down eigenvalues obtained from the self-consistent
Hartree mean-field decoupling are merged into a single set and plotted together. In the first column, we fix $h_z = 0.5$ and vary $\lambda$ from $0.5$ to $0.8$ in panels (a) and (c). In the second column, $\lambda$ is fixed at $0.5$, while $h_z$ is varied from $0.4$ to $0.8$ in panels (b) and (d), respectively.}
    \label{fig:uegnval}
\end{figure}

\begin{figure}[t]
    \centering
   \includegraphics[width=1.0\linewidth]{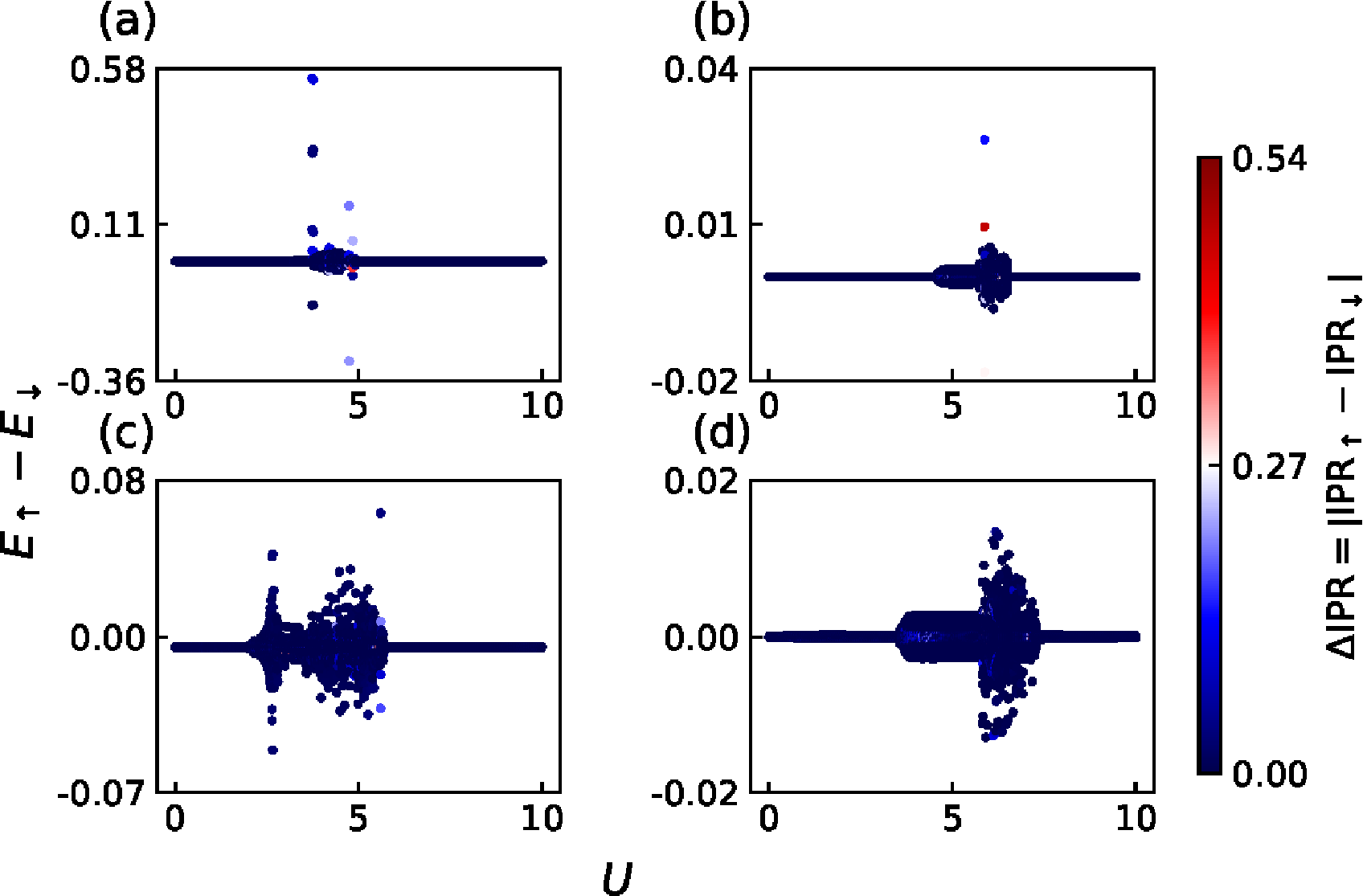}
    \caption{Spin-resolved energy splitting $\Delta E = E_{\uparrow}-E_{\downarrow}$ as a function of $U$, with the color scale representing $|\mathrm{IPR}{\uparrow}-\mathrm{IPR}{\downarrow}|$, shown for the same parameter-space orientation as in the preceding figure, highlighting the onset of interaction-driven spin-dependent localization.
}
    \label{fig:uegndif}
\end{figure}

\begin{figure*}[t]
    \centering
   \includegraphics[width=0.8\linewidth]{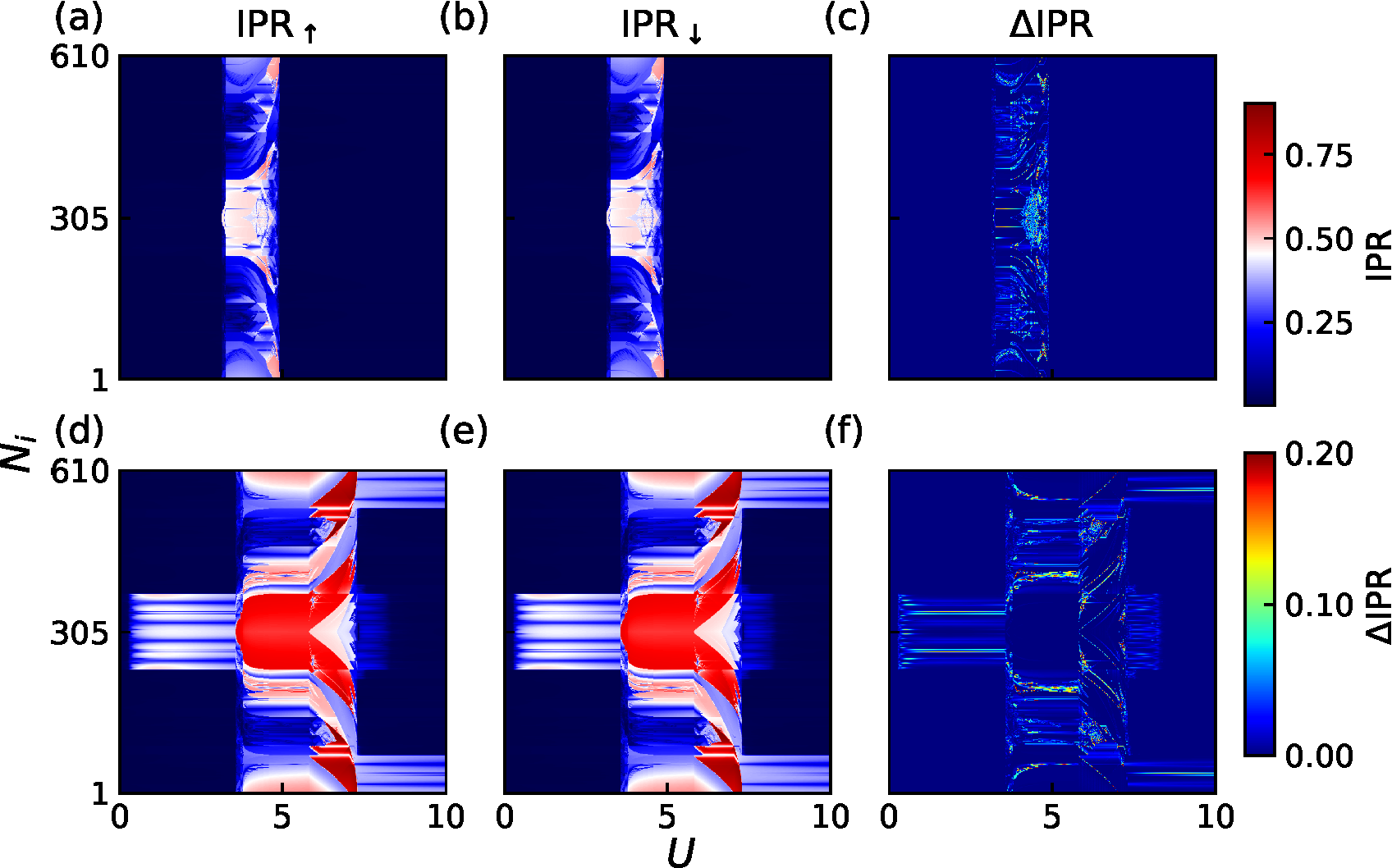}
    \caption{Figure shows the interaction-driven evolution of spin-resolved eigenstate
profiles for $\lambda=0.5$, $h_z=0.4$ (top row) and $\lambda=0.8$, $h_z=0.8$ (bottom row).
Panels (c) and (d) display $\Delta \mathrm{IPR}$, with the color scale capped at $0.20$ for
enhanced visibility, consistent with the IPR-resolved spectral analysis.
}
    \label{fig:ustate}
\end{figure*}

\begin{figure}[t]
    \centering
   \includegraphics[width=1.0\linewidth]{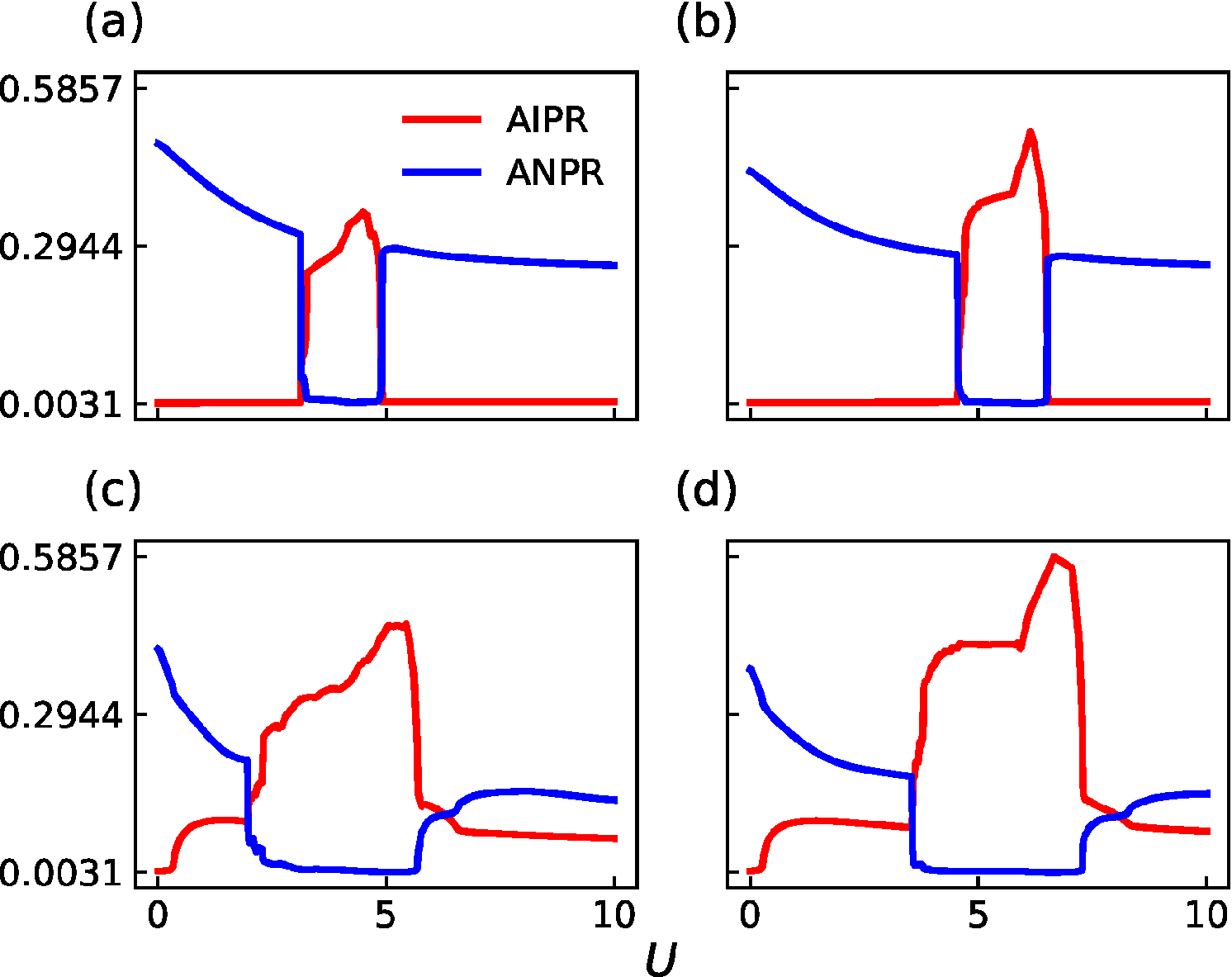}
    \caption{Variation of the average inverse participation ratio (AIPR) and average normalized participation ratio (ANPR) as functions of $U$ for four representative $\lambda$–$h_z$ parameter sets, corresponding to the parameter regimes shown in Fig.~\ref{fig:uegnval}.}
    \label{fig:uainpr}
\end{figure}

\begin{figure}[t]
    \centering
   \includegraphics[width=1.0\linewidth]{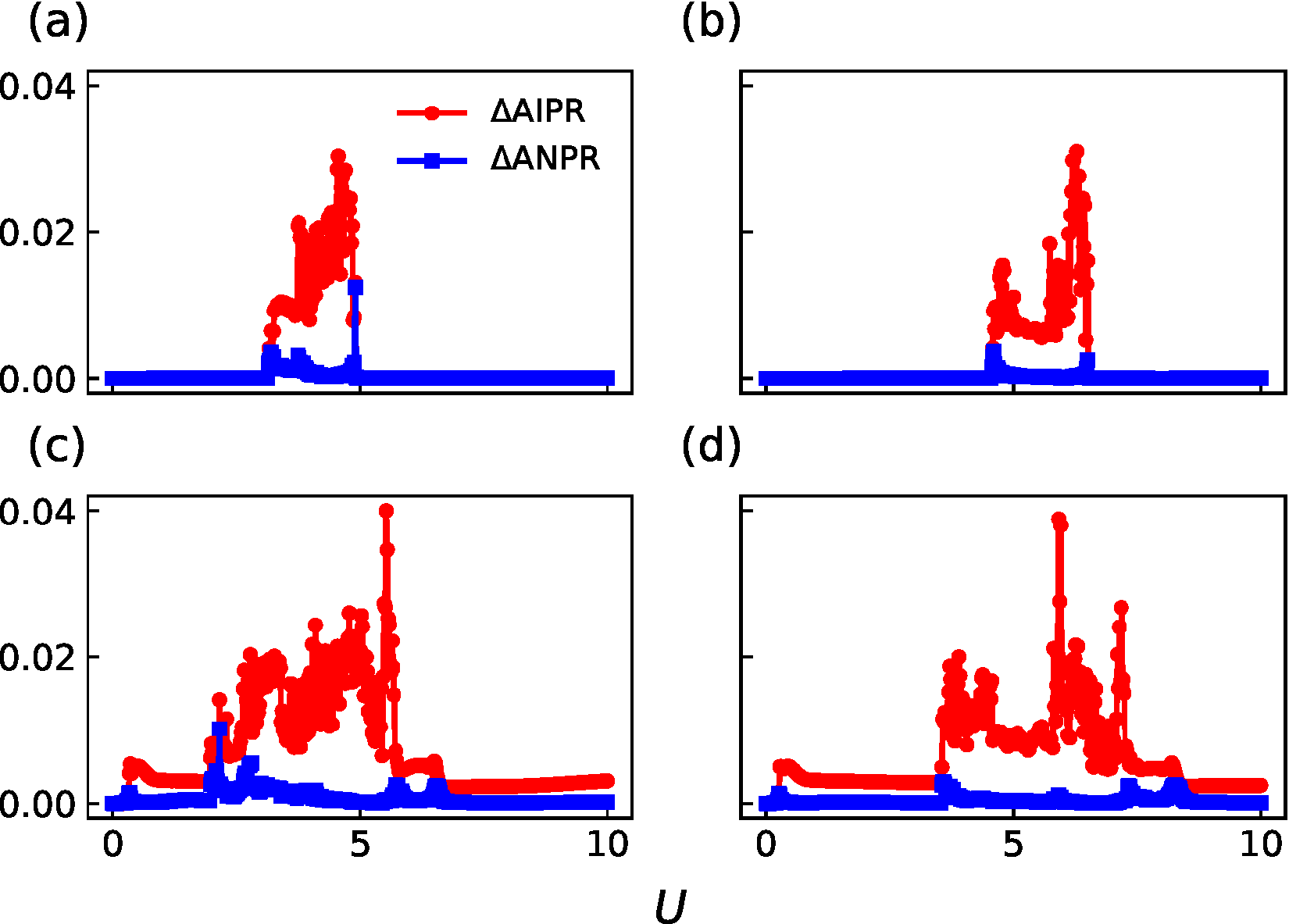}
    \caption{Dependence of the up- and down-spin sector differences in the average inverse participation ratio and average normalized participation ratio ($\triangle$AIPR and $\triangle$ANPR) on $U$ for four representative parameter sets.} 
    \label{fig:udiffainpr}
\end{figure}

\begin{figure}[t]
    \centering
   \includegraphics[width=1.0\linewidth]{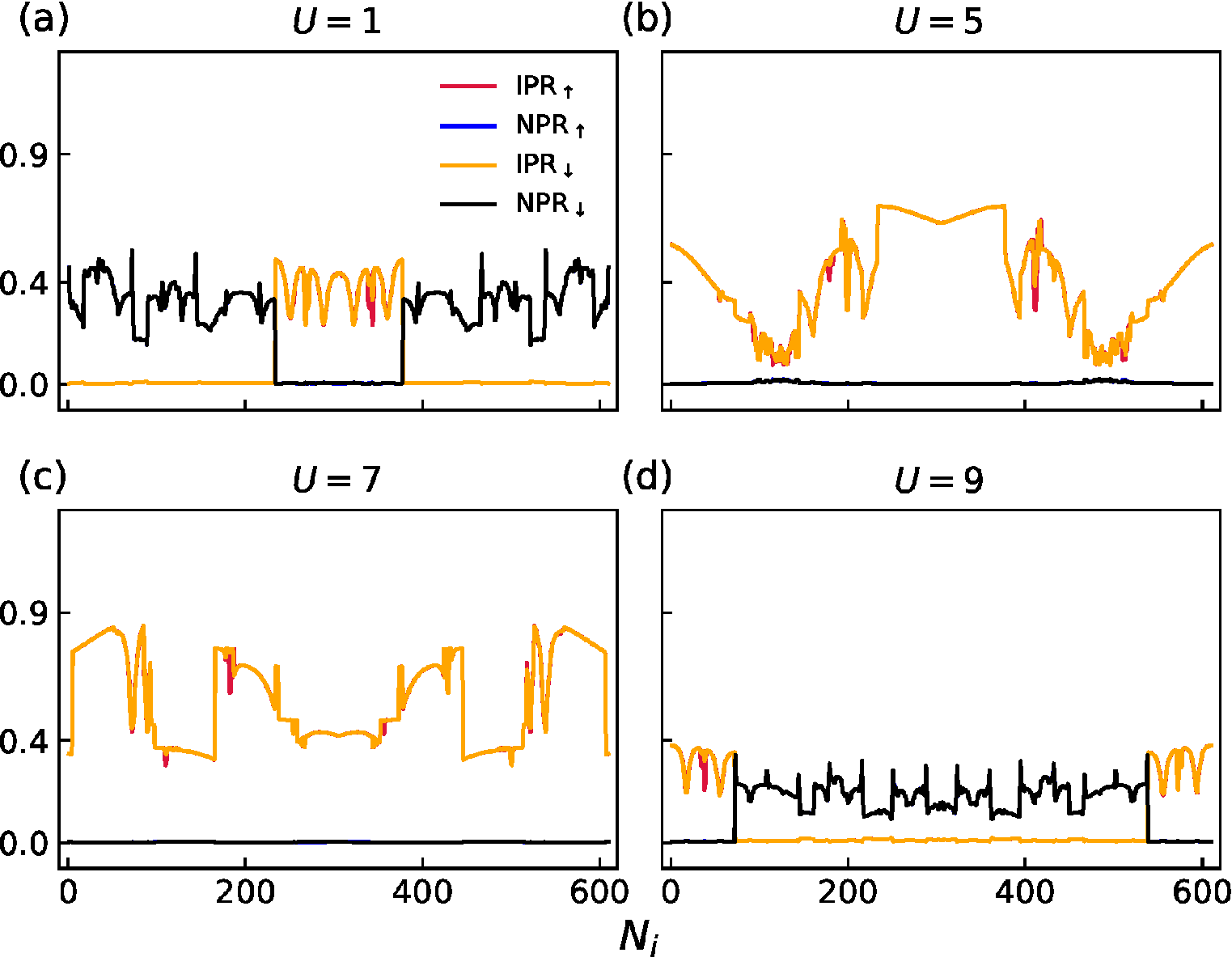}
    \caption{Evolution of spin-resolved localization and delocalization characteristics across the eigenstates, quantified by $IPR_\uparrow$, $IPR_{\downarrow}$, $NPR_\uparrow$, and $NPR_{\downarrow}$, at $\lambda=0.8$ and $h_z=0.8$ for four interaction strengths $U$.}
    \label{fig:iprnprstate1}
\end{figure}

\begin{figure}[t]
    \centering
   \includegraphics[width=1.0\linewidth]{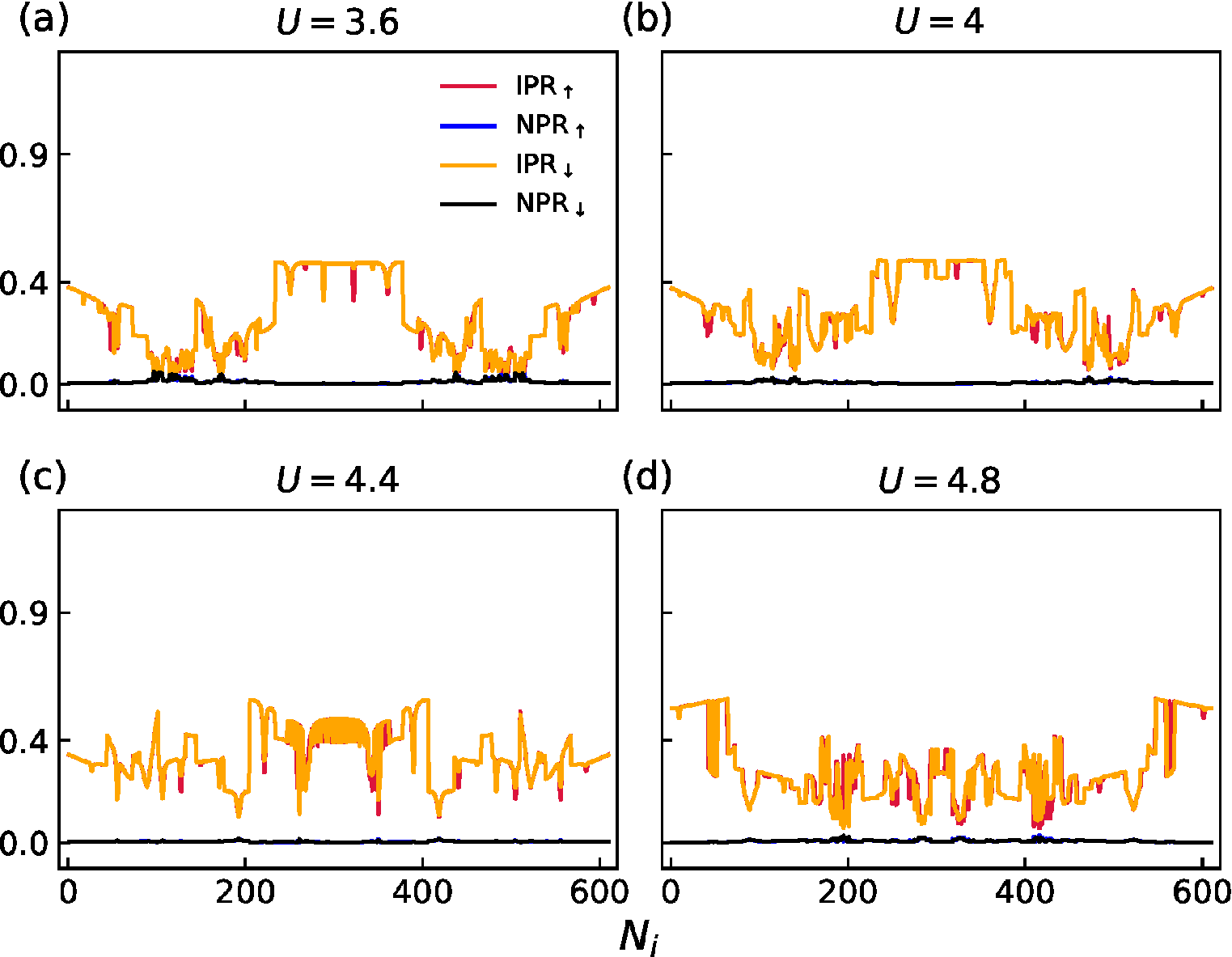}
    \caption{Same as the previous figure, but for $\lambda=0.5$ and $h_z=0.4$.}
    \label{fig:iprnprstate2}
\end{figure}

\section{Result and discussion}
Throughout the calculations, we fix the system size to $610$, corresponding to the largest size that ensures reliable convergence within the mean-field scheme. For phase diagram calculations, however, we restrict the system size to $144$ due to convergence limitations. The hopping parameter is set to $t=1.0$, while all other parameters are specified in the respective sections.
\subsection{Interaction-driven evolution of the IPR-resolved energy spectrum}

In Fig.~\ref{fig:uegnval}, we show the inverse participation ratio (IPR)-resolved energy eigenvalue spectrum as a function of the on-site interaction strength $U$ for representative values of the quasiperiodic modulation $\lambda$ and the Zeeman field $h_z$, as indicated in the caption. For each $U$, all spin-up and spin-down mean-field eigenvalues are combined and plotted, highlighting the interaction-driven restructuring of the effective single-particle spectrum. At $U=0$, all panels exhibit four well-defined subbands arising from the combined effects of spin splitting and lattice modulation. With increasing $U$, interaction-driven correlations progressively reorganize the spectrum, causing the four subbands to merge into two effective bands, indicative of a pronounced reconstruction of the underlying eigenstates.

Focusing on subplot Fig.~\ref{fig:uegnval}(a) ($\lambda=0.5$, $h_z=0.4$), the eigenstates remain predominantly delocalized at weak interaction strengths, as evidenced by the low IPR across the spectrum. Upon increasing $U$, an intermediate interaction window emerges where a pronounced enhancement of IPR is observed, indicating interaction-driven localization. In this regime, additional sets of eigenvalues appear that are absent at lower $U$, forming localized bands, while the previously dispersive subbands also acquire localized character. Interestingly, for sufficiently large $U$, the spectrum reorganizes into two subbands with reduced IPR, indicating a re-entrant delocalized phase driven by strong-coupling correlations.

In subplot Fig.~\ref{fig:uegnval}(b), the Zeeman field strength is increased from $h_z=0.4$ to $h_z=0.8$ while keeping the quasiperiodic modulation fixed at $\lambda=0.5$. The overall behavior remains qualitatively similar to that observed in subplot (a), with the primary difference being a shift of the interaction-driven transition window toward higher values of $U$. Within this window, four well-separated subbands with clearly resolved gaps are visible. Notably, the inner two subbands exhibit enhanced localization compared to subplot (a), indicating that increasing $h_z$ strengthens localization and drives the transition to occur at larger interaction strengths.

A pronounced modification of the spectrum is observed in Fig.~\ref{fig:uegnval}(c) as the quasiperiodic strength is increased from $\lambda=0.5$ to $\lambda=0.8$ at $h_z=0.4$. At low interaction strengths $U$, the inner two subbands out of the four already exhibit localization, a feature that was absent in the weaker quasiperiodic case. As $U$ is increased, the intermediate interaction window in which additional bands emerge becomes broader, and these newly formed bands display enhanced localization compared to subplot (a). At larger $U$, localization becomes spectrally nonuniform: while many eigenstates remain strongly delocalized, a subset displays enhanced localization compared to the lower-$\lambda$ case.

Subplot Fig.~\ref{fig:uegnval}(d) exhibits behavior largely analogous to that of subplot (c), with the key distinction arising from the increased quasiperiodic strength, $\lambda=0.8$, at a fixed Zeeman field $h_z=0.8$. As a result, the interaction-driven transition window is shifted relative to subplot (c). Within this window, clear gaps open between all four subbands, and the inner two subbands display stronger localization than in subplot (c), indicating an enhanced interplay between quasiperiodicity and Zeeman splitting.

To characterize the interaction-induced spin asymmetry, we analyze the spin-resolved energy
splitting $\Delta E = E_{\uparrow}-E_{\downarrow}$ as a function of the Hubbard interaction $U$,
as shown in Fig.~\ref{fig:uegndif}. For each $U$, the energy difference between the corresponding
spin-up and spin-down mean-field eigenvalues is evaluated and color-coded by the difference of
their inverse participation ratios, $|\mathrm{IPR}_{\uparrow}-\mathrm{IPR}_{\downarrow}|$.
This representation directly probes the degree of spin-dependent localization induced by
interactions. While extended eigenstates typically exhibit similar spatial profiles for both
spin channels, resulting in a small IPR difference and a relatively smooth spin splitting,
localized or quasi-localized states respond asymmetrically to the self-consistent Hartree
potential, leading to enhanced spin-dependent localization and larger values of
$|\mathrm{IPR}_{\uparrow}-\mathrm{IPR}_{\downarrow}|$. The combined visualization of
$\Delta E$ and the spin-resolved localization contrast thus highlights the intrinsic coupling
between interaction-driven spectral splitting and spin-selective spatial restructuring in the
interacting quasiperiodic system.

The structure and orientation of the physical parameters in Fig.~\ref{fig:uegndif} are identical
to those in Fig.~\ref{fig:uegnval}. From Fig.~\ref{fig:uegndif}, we observe that the difference
between the spin-up and spin-down eigenvalues remains close to zero up to a certain interaction
strength $U$, the extent of which depends sensitively on the parameter space, such as the
quasiperiodic strength $\lambda$ and the Zeeman field $h_z$. In particular, for
$\lambda=0.5$ and $h_z=0.4$ [subplot (b)], the spin-resolved eigenvalue difference stays
negligibly small up to relatively large values of $U$. More interesting behavior emerges in the intermediate interaction regime, consistent with the
spectral reconstruction observed earlier in Fig.~\ref{fig:uegnval}. In this regime, the
difference between the two spin sectors becomes finite, taking both positive and negative values,
indicating a state-dependent interaction-induced spin splitting. Notably, this region also
exhibits enhanced values of the spin-resolved localization contrast, as reflected by larger
differences in the inverse participation ratios for a subset of eigenstates. At sufficiently
large $U$, the spin-resolved eigenvalue differences as well as the corresponding IPR differences
again tend toward zero, signaling a restoration of effective spin symmetry driven by strong
mean-field renormalization.
\begin{figure}[t]
    \centering
   \includegraphics[width=1.0\linewidth]{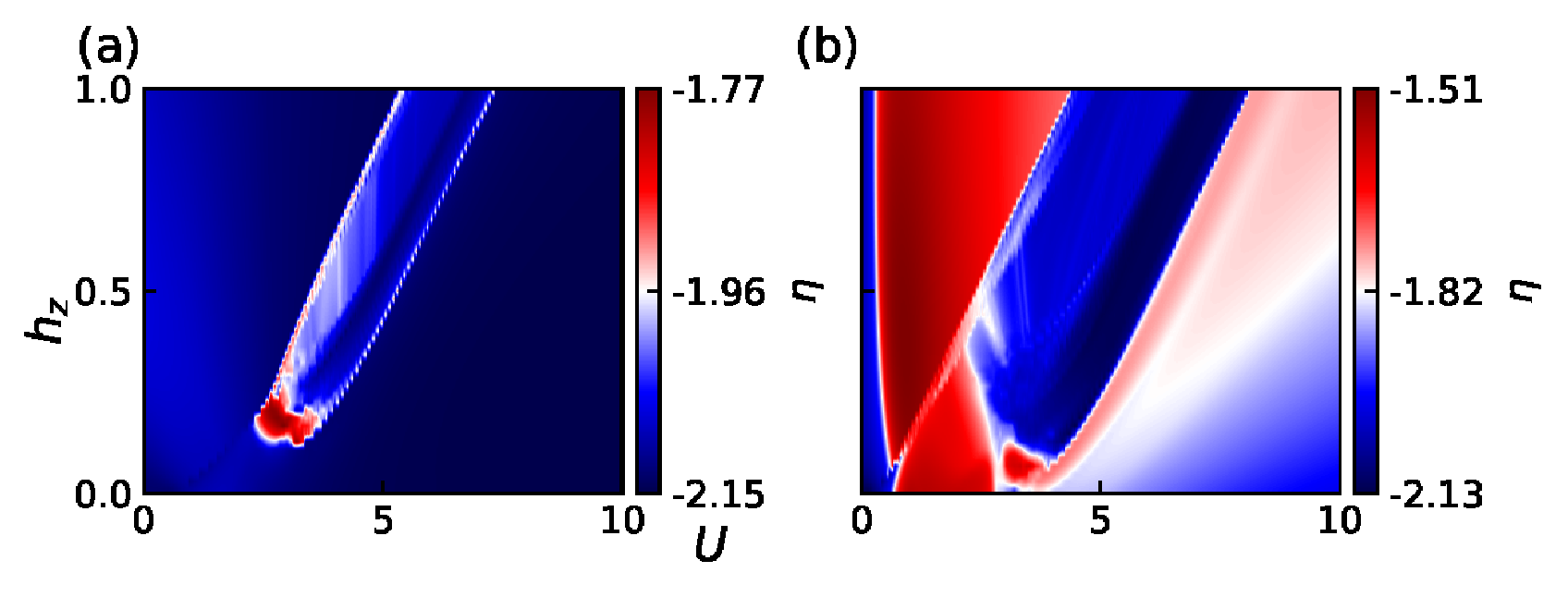}
    \caption{Phase diagrams of $\eta$ in the $h_z$–$U$ parameter space for (a) $\lambda=0.5$ and (b) $\lambda=0.8$ respectively.}
    \label{fig:eta}
\end{figure}

\begin{figure}[t]
    \centering
   \includegraphics[width=1.0\linewidth]{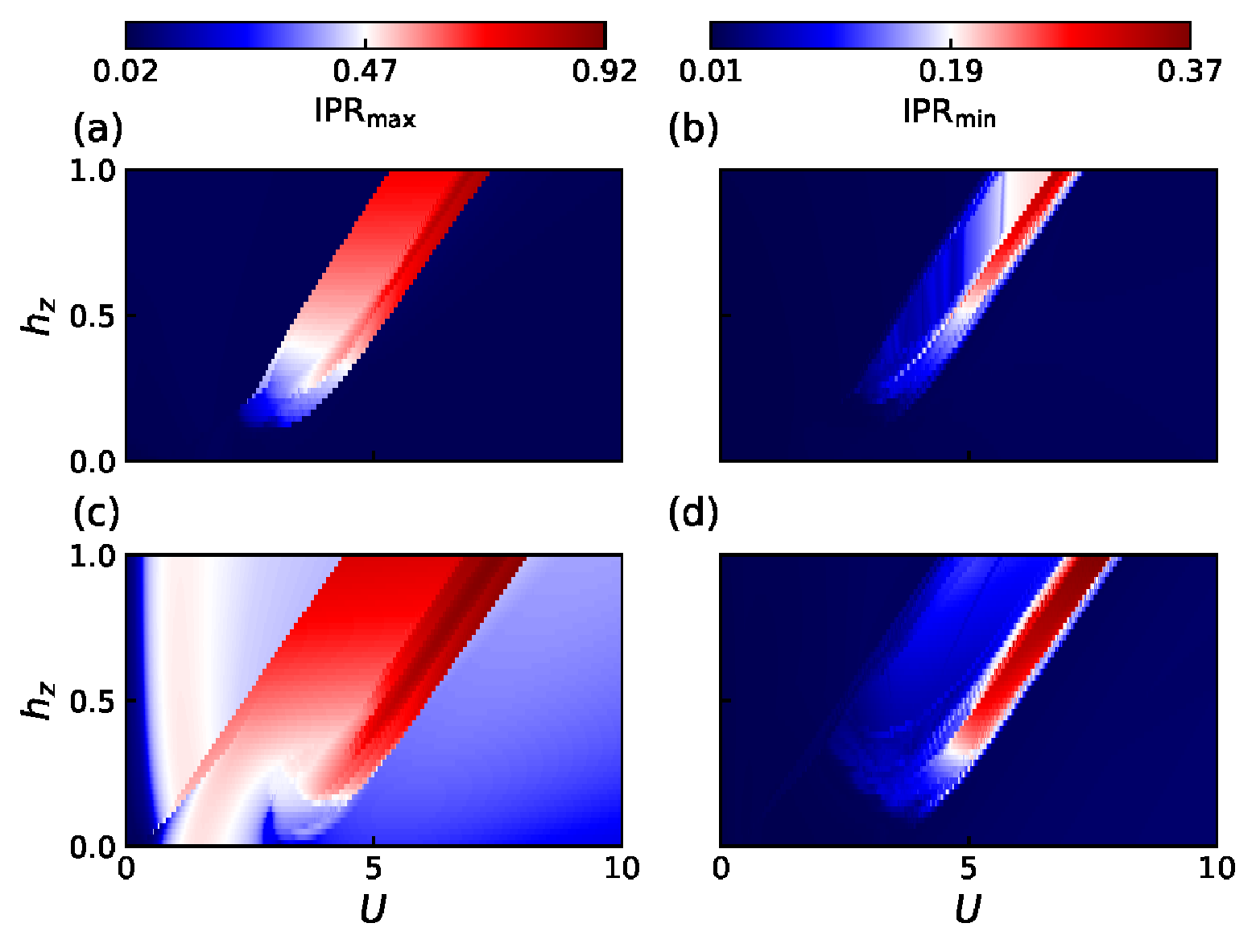}
    \caption{$\mathrm{IPR}_{\mathrm{max}}$ and $\mathrm{IPR}_{\mathrm{min}}$ phase diagrams in the $h_z$–$U$ plane for $\lambda=0.5$ (top) and $\lambda=0.8$ (bottom).}
    \label{fig:extremum}
\end{figure}

\begin{figure}[t]
    \centering
   \includegraphics[width=1.0\linewidth]{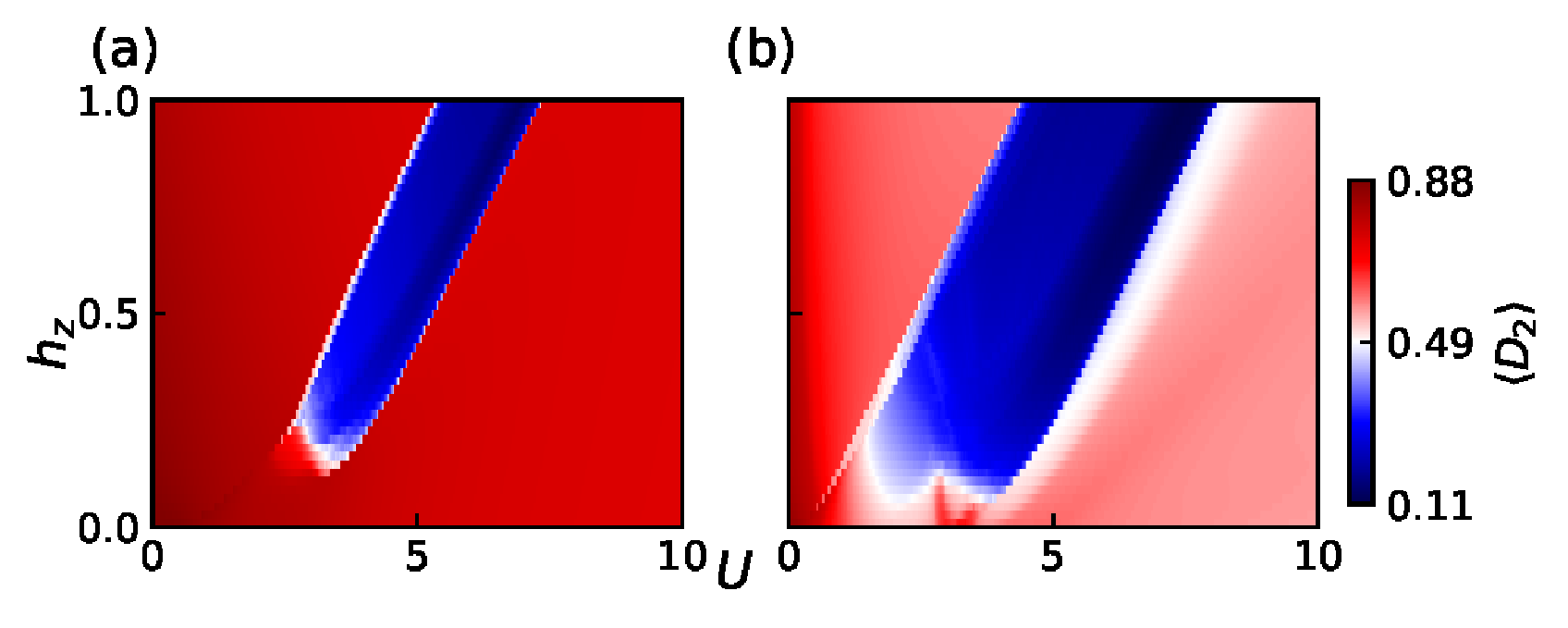}
    \caption{Average fractal dimension $D_2$ in the $h_z$–$U$ plane for (a) $\lambda=0.5$ and (b) $\lambda=0.8$.}
    \label{fig:fractal}
\end{figure}
\subsection{Eigenstate indices with interaction strength $U$}

In Fig.~\ref{fig:ustate}, we illustrate the evolution of the eigenstate profiles as a function of
the interaction strength $U$ for spin-up and spin-down electrons. The top row corresponds to
$\lambda=0.5$ and $h_z=0.4$, while the bottom row shows results for $\lambda=0.8$ and $h_z=0.8$.
In subplots (c) and (d), we present the difference between the spin-resolved inverse participation
ratios, $\Delta \mathrm{IPR} = |\mathrm{IPR}_{\uparrow}-\mathrm{IPR}_{\downarrow}|$. For clarity
of visualization, the color scale for $\Delta \mathrm{IPR}$ is restricted to a maximum value of
$0.20$. The trends observed in the eigenstate profiles are fully consistent with the
IPR-resolved spectral characteristics discussed earlier. In subplots (a), the eigenstates remain predominantly delocalized at weak interaction strengths. As $U$ enters an intermediate regime, a subset of states undergoes pronounced localization,
while at sufficiently large $U$ the eigenstates progressively recover extended character.
Across most of the interaction range, the inverse participation ratios of the spin-up and
spin-down sectors remain nearly identical, indicating minimal spin-dependent localization.
However, in the intermediate interaction window, where significant spectral reconstruction
occurs—the difference between the spin-resolved IPRs becomes finite for a subset of eigenstates.
This enhanced $\Delta\mathrm{IPR}$ (subplot (c)) reflects an interaction-induced spin asymmetry in the spatial structure of the eigenstates and serves as a clear signature of correlation-driven localization
effects. 

Subplots (d) and (e) display the corresponding spin-resolved eigenstate characteristics for
$\lambda=0.8$ and $h_z=0.8$. At weak interaction strengths, the eigenstates near the spectral
edges remain predominantly extended for both spin sectors, while states in the central part
of the spectrum exhibit comparatively stronger localization. As $U$ enters an intermediate
regime (around $U \approx 5$), a pronounced localization emerges in the middle of the spectrum.
Moving away from the spectral center, the states gradually recover extended character; however,
the edge states, which were delocalized at low $U$, begin to show enhanced localization.
Around $U \approx 7$, the edge states become sharply localized. Although qualitatively similar trends are observed in subplots (a) and (b), the localization-
delocalization crossover is significantly more pronounced and clearly resolved in subplots
(d) and (e). At larger interaction strengths, the spectral distribution of localization
contrasts sharply with the weak-$U$ regime: the edge states become more localized than those
near the center of the spectrum. Subplot (f) captures this behavior through the difference
$\Delta\mathrm{IPR}$, which mirrors the same spectral pattern, albeit with a comparatively
smaller magnitude. 

\subsection{$U$-dependence of the average IPR, average NPR, and their spin asymmetries.
}

In Fig.~\ref{fig:uainpr}, we show the evolution of the average inverse participation ratio (AIPR) and the average normalized participation ratio (ANPR) as functions of the interaction strength $U$, with the subplot arrangement following the same parameter-space organization as in the earlier figures. For subplots (a) and (b), where the quasiperiodic strength is fixed at $\lambda=0.5$ and the Zeeman field $h_z$ is increased from $0.4$ to $0.8$, the system remains predominantly extended at weak interactions, undergoes interaction-induced localization in an intermediate-$U$ regime, and subsequently recovers extended behavior at larger $U$, as indicated by enhanced (suppressed) AIPR (ANPR) at intermediate $U$. In contrast, upon increasing $\lambda$ from $0.5$ to $0.8$ while keeping $h_z=0.4$ fixed, the extended phase survives only over a very narrow range of small $U$, followed by rapid localization as $U$ increases, with the system remaining strongly localized throughout the intermediate-$U$ regime. Although partial delocalization tendencies emerge at larger $U$, the system does not fully recover an extended phase, and the AIPR remains finite even at the highest interaction strengths considered.

In Fig.~\ref{fig:udiffainpr}, we analyze the interaction dependence of the spin-resolved differences $\Delta\mathrm{AIPR}$ and $\Delta\mathrm{ANPR}$. At weak interaction strengths, both quantities remain nearly zero, reflecting the fact that spin-up and spin-down electrons experience almost identical effective mean-field potentials and thus exhibit similar spatial distributions. In this regime, the eigenstates are predominantly extended, and the influence of the interaction-induced Hartree term on spin differentiation is minimal.  As $U$ is increased into the intermediate interaction regime, both $\Delta\mathrm{AIPR}$ and $\Delta\mathrm{ANPR}$ develop pronounced maxima. This enhancement coincides with the onset of interaction-driven localization observed in Fig.~\ref{fig:uainpr}, indicating that localized eigenstates are significantly more sensitive to spin-dependent fluctuations of the mean-field potential. In this regime, small differences in the local spin densities are amplified by the Hubbard interaction, leading to an imbalance in the spatial confinement of spin-up and spin-down states. Consequently, localization acts as a magnifying mechanism for spin asymmetry, even though the overall magnitude of the difference remains small. Upon further increasing $U$, both $\Delta\mathrm{AIPR}$ and $\Delta\mathrm{ANPR}$ gradually decrease and approach zero again. This behavior reflects the restoration of a more homogeneous mean-field environment at strong coupling, where interaction effects dominate over quasiperiodic modulation and suppress spin-dependent spatial differentiation. The nonmonotonic dependence of $\Delta\mathrm{AIPR}$ and $\Delta\mathrm{ANPR}$ on $U$ therefore establishes a direct connection between correlation-induced localization and spin-resolved spatial asymmetry in the interacting quasiperiodic system.
\begin{figure*}[!htbp]
\centering
\includegraphics[width=0.98\textwidth]{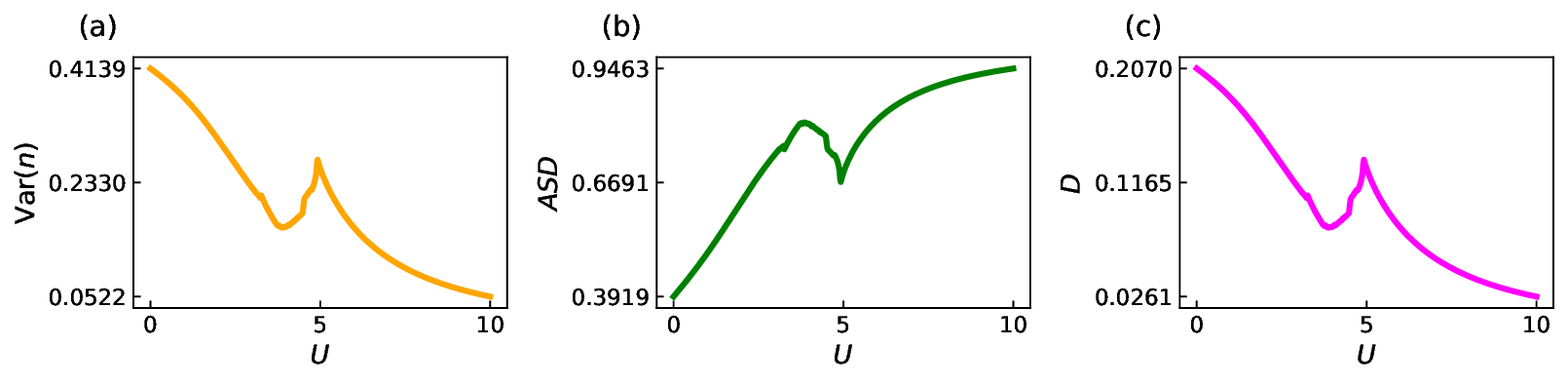}

\vspace{-3mm}

\includegraphics[width=0.98\textwidth]{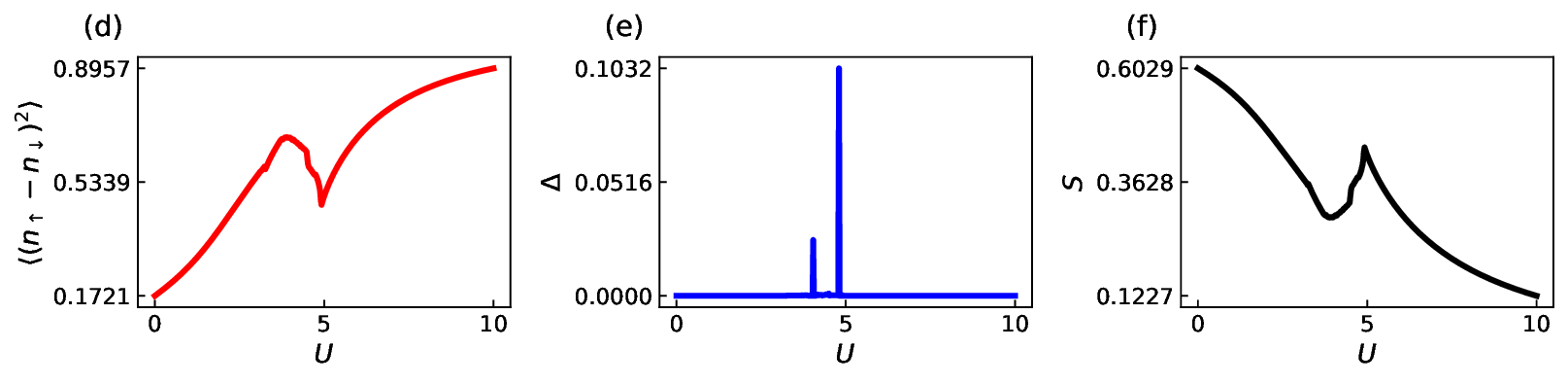}

\caption{(Color online). Dependence of various mean-field observables on $U$: variance (a), ASD (b), double occupancy $D$ (c), $\langle (n_{\uparrow}-n_{\downarrow})^2\rangle$ (d), $\triangle$ (e), and local entropy $S$ (f), shown for $\lambda=0.5$ and $h_z=0.4$.}
\label{fig:nptitall}
\end{figure*}
\subsection{Variation of IPR (NPR) with eigenstates}

In Fig.~\ref{fig:iprnprstate1}, we present the eigenstate-resolved inverse participation ratio (IPR) and normalized participation ratio (NPR) for $\lambda=0.8$ and $h_z=0.8$ at four representative interaction strengths. For weak interaction ($U=1$), the spectrum is predominantly extended, as reflected by low IPR and high NPR values, except for states near the center of the spectrum that exhibit enhanced localization. Upon increasing the interaction to $U=5$, interaction-induced correlations substantially reshape the spatial character of the eigenstates: the system displays a mixed regime in which localized states emerge in the central part of the spectrum, while neighboring states show partial delocalization. This nonuniform localization pattern persists and becomes more pronounced at $U=7$, where a larger fraction of eigenstates exhibits stronger spatial confinement. At strong coupling ($U=9$), the localization profile undergoes a qualitative inversion: the edge states become strongly localized, whereas a broad portion of the spectrum recovers extended character. The spectrum retains a mirror-symmetric structure about its center for all interaction strengths, reflecting the underlying lattice and spin symmetries. Overall, these results demonstrate a nonmonotonic and spectrally selective evolution of localization driven by the interplay between quasiperiodicity, Zeeman splitting, and interaction effects. 

Figure~\ref{fig:iprnprstate2} presents the same set of four subplots for $\lambda=0.5$ and $h_z=0.4$. In this case, a pronounced mirror-symmetric structure of the eigenstate profiles is observed across all interaction strengths. Unlike the higher-$\lambda$ case discussed earlier, the spectral inversion of localization characteristics, previously seen at large $U$ is absent here, indicating a weaker interaction-driven redistribution of localized and extended states in this parameter regime.

We further observe that for most eigenstates the IPR (NPR) of the spin-up and spin-down sectors overlap almost exactly, indicating nearly identical localization properties for the two spin channels. However, for a subset of eigenstates this overlap breaks down, manifested as distinct deviations that appear, for instance, as pronounced spikes (red color) in the IPR spectrum. These features signal eigenstates for which the spatial localization characteristics of the two spin components differ quantitatively, reflecting a state-dependent spin asymmetry induced by interactions. Such deviations, signaling state-dependent spin asymmetry in localization, are more prominent in Fig.~\ref{fig:iprnprstate2} than in Fig.~\ref{fig:iprnprstate1}.

\subsection{Phase diagram of $\eta$, the extrema of the IPR, and the fractal dimension $D_2$}

In Fig.~\ref{fig:eta}, we present the phase diagram of $\eta$ over the full parameter space, with $h_z$ varied along the $y$ axis and $U$ along the $x$ axis. Such $\eta$-based phase diagrams are commonly used to identify localized, delocalized, and intermediate regimes. For numerical simplicity and to ensure reliable convergence, we restrict our analysis to a system size of $L=144$, which is sufficient to capture the essential characteristics of the different phases. Subplots (a) and (b) correspond to two representative quasiperiodic strengths, $\lambda=0.5$ and $\lambda=0.8$, respectively.
Since $\eta$ alone cannot distinguish between fully delocalized and fully localized phases, we additionally analyze the extrema of the IPR, namely its maximum and minimum values, as shown in Fig.~\ref{fig:extremum}. The first and second rows correspond to the lower and higher values of $\lambda$, respectively, and closely follow the structure of the $\eta$ phase diagram. From Fig.~\ref{fig:eta}(a), a strip-like region emerges beyond a critical value of $h_z$, which gradually broadens with increasing $h_z$. A detailed inspection of the corresponding maximum and minimum IPR maps in Fig.~\ref{fig:extremum}(a) and (b) reveals that the regions on both sides of this strip correspond to fully delocalized phases. Within the strip, the red-colored region in Fig.~\ref{fig:extremum}(b) signifies a fully localized phase. In contrast, the off-white to light bluish regions in Fig.~\ref{fig:extremum}(b), along with the corresponding regions in Fig.~\ref{fig:extremum}(a) and Fig.~\ref{fig:eta}(a), can be identified as an intermediate phase, characterized by finite minimum IPR values that do not approach zero.

For $\lambda=0.8$, as shown in Fig.~\ref{fig:eta}(b) and the corresponding Figs.~\ref{fig:extremum}(c) and (d), the fully delocalized regime survives only within a very narrow window of $U$, indicated by the dark blue region in Fig.~\ref{fig:extremum}(c). A closer inspection of the minimum IPR map reveals a red strip that is considerably thicker than in the $\lambda=0.5$ case, signifying an extended localized region. In contrast, the off-white to light bluish regions in Fig.~\ref{fig:extremum}(d), which are noticeably narrower than in the lower-$\lambda$ case, can be identified as an intermediate phase. An important distinction emerging from Figs.~\ref{fig:extremum}(c) and (d) is that, although the minimum IPR approaches zero (dark blue region in subplot (d)), the maximum IPR remains finite, indicating the coexistence of states with small and moderate IPR values. Consequently, upon averaging over all states, the system effectively enters a localized phase, a behavior that is qualitatively different from the $\lambda=0.5$ case, where both the minimum and maximum IPR values on either side of the strip remain close to zero, reflecting a uniformly extended phase.

In Fig.~\ref{fig:fractal}, we present the phase diagram of the average fractal dimension $\langle D_2\rangle$ in the $h_z$–$U$ plane for $\lambda=0.5$ and $0.8$, shown in subplots (a) and (b), respectively. For $\lambda=0.5$ [subplot (a)], a reddish region with $\langle D_2\rangle \approx 0.88$ clearly indicates a delocalized phase, while the dark blue region, where $\langle D_2\rangle$ is reduced to values around $0.11$, corresponds to a localized window. The intermediate off-white to light bluish regions, characterized by $\langle D_2\rangle \gtrsim 0.3$, can therefore be identified as a critical (multifractal) regime. For the higher quasiperiodic strength $\lambda=0.8$ [subplot (b)], a fully extended phase with large $\langle D_2\rangle$ survives only within a very narrow window of $U$, consistent with our earlier analysis. In contrast, a broad region with intermediate values $\langle D_2\rangle \sim 0.7$ signals an enhanced critical regime, while the dark bluish region expands significantly, indicating a strongly localized phase, in agreement with the corresponding IPR-extrema phase diagrams. Finally, we note that the mean-field calculations are highly sensitive to system size for achieving convergence; therefore, we restrict our analysis to $L=144$ lattice sites, which is sufficient to capture the essential macroscopic features and provide a qualitative overview of the phase structure.

\subsection{Different mean field parameters with $U$}
To further quantify the interaction-driven restructuring of the electronic states,
we examine the variance of the local site occupations~\cite{mf1}, defined as
\begin{equation}
\mathrm{Var}_{\mathrm{sites}}
= \langle n_{\uparrow} \rangle \left( 1 - \langle n_{\uparrow} \rangle \right)
+ \langle n_{\downarrow} \rangle \left( 1 - \langle n_{\downarrow} \rangle \right),
\end{equation}
where $\langle n_{\sigma} \rangle$ denotes the spin-resolved mean density obtained from
the self-consistent Hartree solution.

In Fig.~\ref{fig:nptitall}(a), we present the interaction dependence of the density fluctuation
$\mathrm{Var}(n)$, obtained by averaging $\mathrm{Var}_{\mathrm{sites}}$ over all lattice sites.
In the weakly interacting regime, $\mathrm{Var}(n)$ assumes relatively large values, reflecting
nearly homogeneous charge distributions associated with extended eigenstates; correspondingly,
the average inverse participation ratio (AIPR) remains low. As the Hubbard interaction $U$ is
increased, $\mathrm{Var}(n)$ initially decreases, following a trend opposite to that of AIPR.
However, upon entering an intermediate interaction window, a subset of eigenstates undergoes
localization driven by the combined effects of quasiperiodicity and the self-consistent Hartree
potential. This interaction-induced localization gives rise to pronounced spatial density
modulations, enhancing site-to-site charge fluctuations and leading to a noticeable increase in
$\mathrm{Var}(n)$. The simultaneous enhancement of $\mathrm{Var}(n)$ and AIPR in this regime
establishes a direct correspondence between density inhomogeneity and eigenstate localization.
At larger values of $U$, the effective mean-field potential becomes smoother again, and the system
returns to a weakly inhomogeneous or extended regime, where both $\mathrm{Var}(n)$ and AIPR are
suppressed. The resulting nonmonotonic evolution of $\mathrm{Var}(n)$ with $U$ therefore closely
tracks the behavior of AIPR, demonstrating that density fluctuations provide an independent and
complementary probe of interaction-driven localization in the quasiperiodic Hubbard model.

In Fig.~\ref{fig:nptitall}(b) we plot the average spin-density imbalance (ASD)~\cite{mf2},
defined as
\begin{equation}
\mathrm{ASD}=\frac{1}{L}\sum_{i=1}^{L}
\big| \langle n_{i,\uparrow}\rangle-\langle n_{i,\downarrow}\rangle \big| ,
\end{equation}
which directly quantifies the site-averaged spin polarization in the interacting
system. At weak interaction strengths, the ASD increases with $U$, reflecting the
enhancement of spin imbalance induced by the Zeeman field $h_z$ in the presence of
a nearly homogeneous Hartree potential. As $U$ is increased into the intermediate
regime, the combined effects of quasiperiodicity and interaction-driven Hartree
fields lead to partial localization and strong spatial inhomogeneity of the local
densities. In this regime, the spin polarization becomes highly nonuniform across
the lattice, resulting in a partial cancellation of local spin imbalances when
averaged over sites and giving rise to a pronounced dip in the ASD. Upon further
increasing $U$, the interaction term dominates and enforces a more robust local
spin polarization throughout the system, thereby increasing the ASD again. The
resulting nonmonotonic dependence of the ASD on $U$ thus captures the competition
between Zeeman-induced spin splitting and interaction-driven localization, and
provides a complementary spin-resolved signature of the interaction-controlled
restructuring of the electronic states. A similar pattern and characteristic behavior are also observed in Fig.~\ref{fig:nptitall}(d).

In Fig.~\ref{fig:nptitall}(c) we plot the interaction dependence of the site-averaged quantity
\begin{equation}
D=\frac{1}{L}\sum_{i=1}^{L}
\langle n_{i,\uparrow}\rangle \langle n_{i,\downarrow}\rangle ,
\end{equation}
which represents the mean-field probability of simultaneous spin-up and spin-down
occupation at the same lattice site. In the weakly interacting regime, $D$
decreases gradually with increasing $U$ due to the repulsive Hubbard interaction,
which energetically disfavors local double occupancy. As $U$ enters an intermediate
regime, the combined effects of quasiperiodicity and interaction-induced Hartree
potentials drive partial localization and strong spatial modulation of the local
densities. In this regime, electrons become concentrated on a subset of lattice
sites, enhancing the overlap between $\langle n_{i,\uparrow}\rangle$ and
$\langle n_{i,\downarrow}\rangle$ on those sites and leading to a noticeable
increase in $D$. Upon further increasing $U$, the repulsive interaction dominates
over density inhomogeneity, strongly suppressing simultaneous spin occupancy across
the lattice and causing $D$ to decrease again. The resulting nonmonotonic
behavior of $D$ therefore reflects a competition between interaction-driven
localization, which enhances local density overlap, and strong on-site repulsion,
which suppresses double occupancy. Its close correspondence with the behavior of
$\mathrm{Var}(n)$ further establishes $D$ as an independent real-space indicator
of interaction-induced localization within the Hartree mean-field framework.
\begin{figure}[t]
    \centering
   \includegraphics[width=1.0\linewidth]{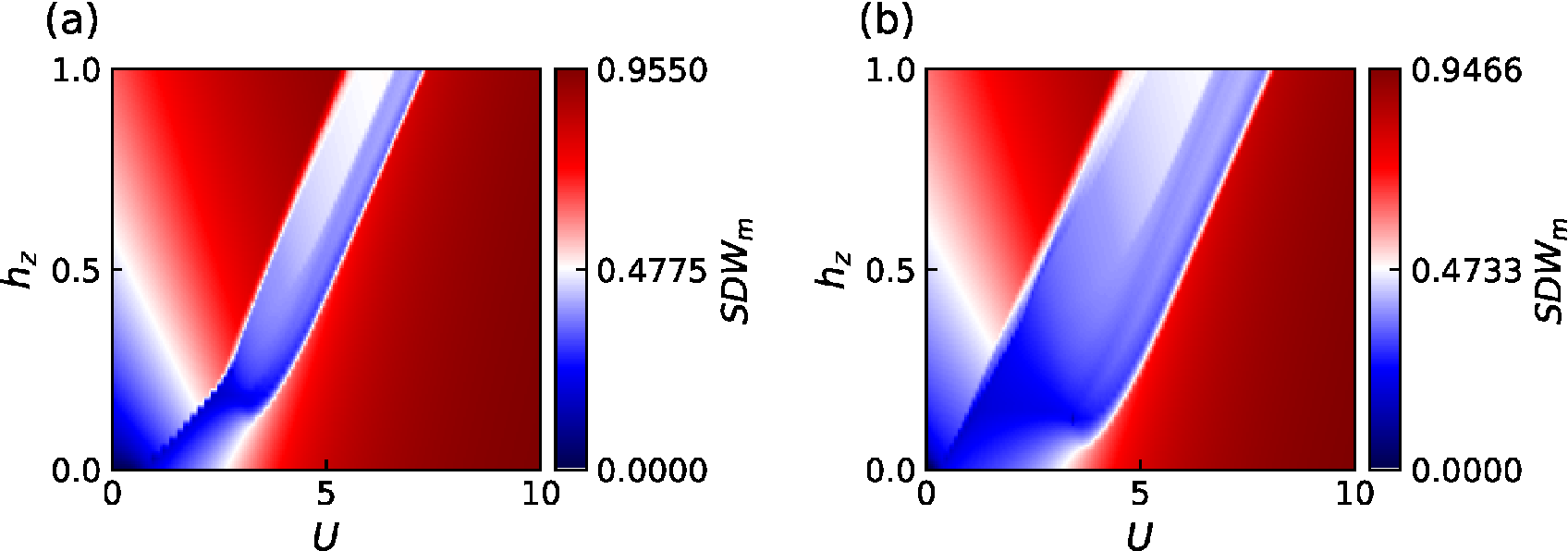}
    \caption{Variation of SDW in the $h_z$–$U$ plane for $144$ sites. Panels (a) and (b) correspond to $\lambda=0.5$ and $0.8$, respectively.}
    \label{fig:sdw}
\end{figure}
In Fig.~\ref{fig:nptitall}(e) we present the interaction dependence of the single-particle excitation gap
\begin{equation}
\Delta=E_{\mathrm{unocc}}-E_{\mathrm{occ}},
\end{equation}
where $E_{\mathrm{occ}}$ and $E_{\mathrm{unocc}}$ denote, respectively, the highest occupied and
lowest unoccupied eigenvalues obtained from the combined spin-up and spin-down mean-field
spectrum. In the weakly interacting regime, the gap remains nearly zero, consistent with an
extended phase characterized by delocalized eigenstates and a quasi-continuous spectrum.
As the Hubbard interaction $U$ is increased into an intermediate regime, a finite gap
develops, signaling a substantial restructuring of the effective single-particle spectrum.
This gap opening coincides with the enhancement of AIPR, $\mathrm{Var}(n)$, and the site-averaged
overlap $D$, indicating that interaction-induced localization and strong density
inhomogeneity lead to an energetic separation between occupied and unoccupied states.
In this regime, the self-consistent Hartree potential generates site-dependent energy shifts,
which lift the near-degeneracies present at low $U$ and stabilize a gapped spectrum.
Upon further increasing $U$, the gap decreases again and tends toward smaller values,
reflecting a partial restoration of spectral continuity associated with the re-emergence of
extended or weakly localized eigenstates. The nonmonotonic behavior of $\Delta$ therefore
mirrors that of AIPR and density-fluctuation measures, establishing the excitation gap as an
energetic manifestation of interaction-driven localization in the quasiperiodic Hubbard model.

We further investigate the interaction-driven evolution of the local electronic structure
through the site-averaged local entropy, defined as
\begin{equation}
S_{\mathrm{loc}}
= -\left[n_{\uparrow}\ln(n_{\uparrow}) + n_{\downarrow}\ln(n_{\downarrow})\right],
\end{equation}
with $n_{\uparrow}$ and $n_{\downarrow}$ obtained from the self-consistent Hartree solution.
Figure~\ref{fig:nptitall}(f) displays the spatially averaged quantity $\langle S_{\mathrm{loc}}\rangle$, denoted by $S$, as a function of the interaction strength $U$.
At weak coupling, $S$ assumes relatively large values,
reflecting nearly homogeneous spin densities and a high degree of local configurational
uncertainty characteristic of extended eigenstates. As the interaction strength $U$ is increased into an intermediate regime, the system undergoes
a qualitative reorganization driven by the interplay of quasiperiodicity and
interaction-induced Hartree fields. In this regime, strong site-dependent density
modulations develop, leading to pronounced spatial variations of both $n_{\uparrow}$ and
$n_{\downarrow}$. These inhomogeneities enhance local charge and spin fluctuations rather than suppressing them, resulting in a marked increase of the site-averaged local entropy
$S$. The entropy peak coincides with the maxima observed in
$\mathrm{Var}(n)$ and $D$, as well as with enhanced AIPR and the emergence of a finite
single-particle gap, establishing the intermediate-$U$ window as the regime of strongest
correlation effects. Upon further increasing $U$, the effective mean-field potential becomes
smoother and the system crosses over into a weakly inhomogeneous or re-extended regime,
where density fluctuations are reduced. Consequently, $S$ decreases again, mirroring the behavior of $\mathrm{Var}(n)$ and $D$. The resulting
nonmonotonic dependence of the local entropy on $U$ therefore provides an entropic
fingerprint of interaction-driven localization and spectral reconstruction in the
quasiperiodic Hubbard model.

\subsection{Variation of SDW phase diagram }
The spin-density-wave (SDW)~\cite{mf2} amplitude is obtained from the converged
spin-resolved local densities $\langle n_{i,\uparrow} \rangle$ and
$\langle n_{i,\downarrow} \rangle$ as
\begin{equation}
m(\mathbf{q})
=
\frac{1}{L}
\left|
\sum_{i=1}^{L}
e^{ i \mathbf{q}\cdot i }
\left(
\langle n_{i,\uparrow} \rangle
-
\langle n_{i,\downarrow} \rangle
\right)
\right|,
\end{equation}
where $L$ denotes the total number of lattice sites and $\mathbf{q}$
is the SDW ordering wave vector.
The maximum SDW response is then defined as
\begin{equation}
SDW_{m}
=
\max_{\mathbf{q}}
\, m(\mathbf{q}),
\end{equation}
which is evaluated over all allowed wave vectors in the Brillouin zone. 

In Fig.~\ref{fig:sdw}(a), we present the phase diagram of the maximum spin–density–wave amplitude $\mathrm{SDW}_{m}$ in the $h_z$–$U$ parameter space for $\lambda=0.5$. At $U=0$, $\mathrm{SDW}_{m}$ remains vanishingly small up to a finite threshold value of the staggered field $h_z$, beyond which it increases monotonically and reaches values close to saturation ($\mathrm{SDW}_{m}\simeq 0.95$) at $h_z=1$. A similar behavior is observed along the interaction axis: for $h_z=0$, $\mathrm{SDW}_{m}$ is initially zero and becomes finite only beyond a critical interaction strength, increasing steadily with $U$. A more intriguing behavior emerges at finite $h_z$, where $\mathrm{SDW}_{m}$ is already nonzero at weak interaction strengths but decreases upon increasing $U$, reaching a minimum at intermediate $U$. This suppression of SDW order coincides with the interaction regime in which pronounced spectral reconstruction and localization effects were identified in earlier figures. Upon further increasing $U$, the SDW amplitude recovers and grows again, leading to a nonmonotonic high–low–high evolution of $\mathrm{SDW}_{m}$. This re-entrant behavior of the SDW order parameter is closely correlated with the extended–localized–extended crossover observed in the localization diagnostics, highlighting a strong interplay between interaction-driven magnetic ordering and spatial localization of the electronic states.

Figure~\ref{fig:sdw}(b), the quasiperiodic modulation strength is increased to $\lambda=0.8$. As a result, the intermediate interaction regime is significantly enlarged, consistent with our earlier observations. In particular, the nonmonotonic high–low–high evolution of the maximum SDW amplitude becomes more pronounced: the intermediate region where $\mathrm{SDW}_{m}$ is suppressed broadens, while the low- and high-$U$ regimes with enhanced SDW order are correspondingly reduced. This expansion of the intermediate low-$\mathrm{SDW}_{m}$ region reflects the strengthened influence of quasiperiodicity, which promotes localization and suppresses coherent spin modulation over a wider range of interaction strengths. The resulting behavior further supports the close connection between the re-entrant SDW response and the extended–localized–extended crossover identified in the localization and spectral analyses.

\subsection{Real time dynamics}
 \begin{figure}[t]
    \centering
   \includegraphics[width=1.0\linewidth]{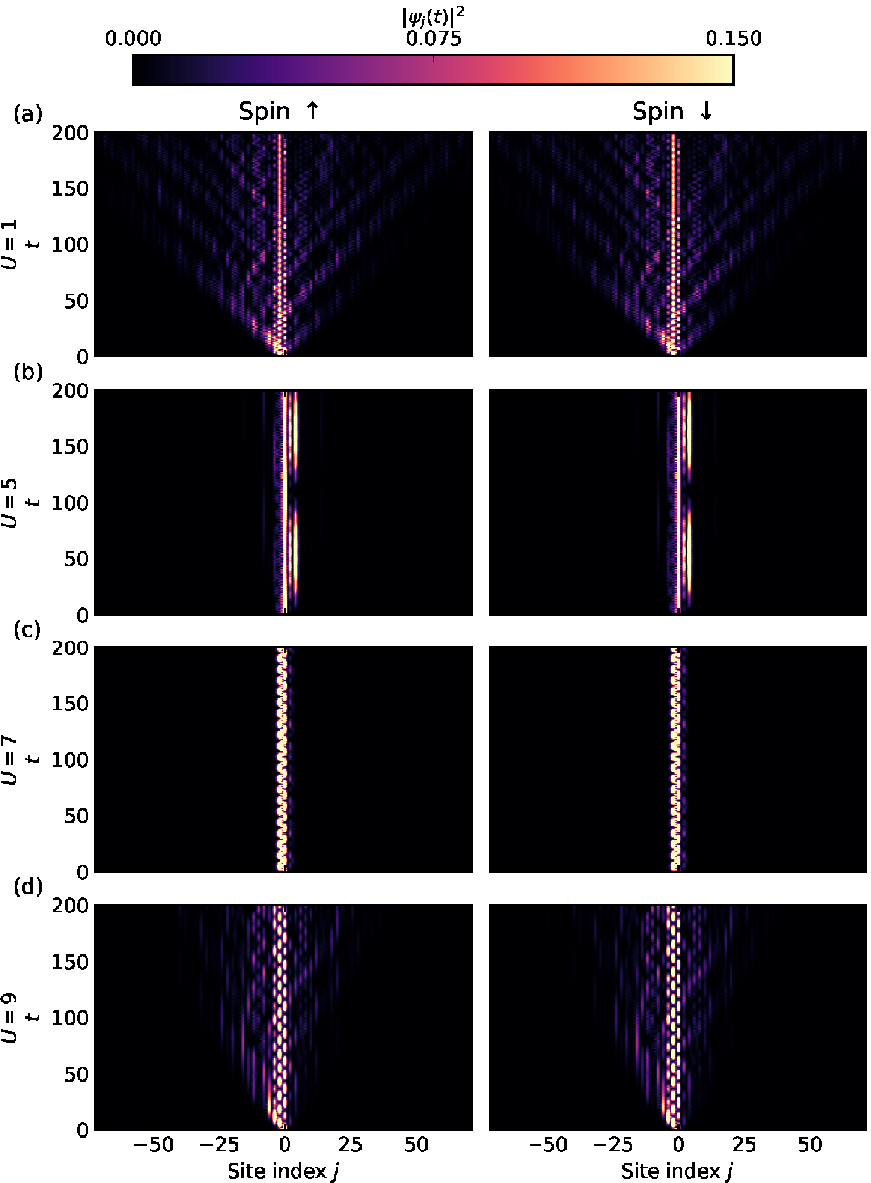}
    \caption{Spatiotemporal evolution of the probability density $|\psi(t)|^2$ across the lattice, resolved for both spin sectors. Panels (a)–(d) illustrate the extended, intermediate, localized, and intermediate dynamical regimes obtained at interaction strengths $U=1$, $5$, $7$, and $9$, respectively, for $\lambda=0.8$ and $h_z=0.8$. The system size is fixed at $L=144$ to enhance the visibility of the time evolution and to clearly distinguish extended from localized propagation in both the up- and down-spin channels.}
    \label{fig:dyna1}
\end{figure}
The real-time evolution~\cite{dyna1,dyna2} of a spin-resolved single-particle wave packet governed by the effective mean-field Hamiltonian. To probe the dynamical transport characteristics of the system, we consider an initially localized excitation prepared at a specific lattice site,
\[
\lvert \psi_{\sigma}(0) \rangle = \lvert j_{0}, \sigma \rangle ,
\]
where $j_{0}$ denotes the initial site and $\sigma=\uparrow,\downarrow$ labels the spin degree of freedom. The subsequent time evolution of the wave packet is dictated by the unitary time-evolution operator,
\[
\lvert \psi_{\sigma}(t) \rangle = e^{-i H_{\sigma} t}\, \lvert \psi_{\sigma}(0) \rangle ,
\]
where we set $\hbar=1$ and $H_{\sigma}$ represents the spin-dependent effective Hamiltonian obtained from the self-consistent Hartree decoupling. Since the dynamics is governed by an effective single-particle description, the time-evolved state can be expressed in the lattice basis as
\[
\lvert \psi_{\sigma}(t) \rangle = \sum_{j=1}^{L} \psi_{j,\sigma}(t)\, \lvert j, \sigma \rangle ,
\]
with $\psi_{j,\sigma}(t)$ denoting the probability amplitude of finding a particle with spin $\sigma$ at site $j$ and time $t$. The spatiotemporal spreading of the wave packet is therefore quantified through the site- and spin-resolved probability density $|\psi_{j,\sigma}(t)|^{2}$.

Figure~\ref{fig:dyna1} displays the real-time evolution of the spin-resolved probability density $|\psi_{j,\sigma}(t)|^{2}$ as a function of the site index $j$ and time $t$ for both spin sectors, evaluated at four representative values of the interaction strength $U$ for $\lambda=0.8$ and $h_z=0.8$. A reduced system size of $L=144$ is chosen to clearly resolve the dynamical features. Consistent with the static spectral and localization diagnostics discussed earlier, the spin-up and spin-down components exhibit nearly identical dynamical signatures across all interaction regimes. At weak interaction ($U=1$), the initially localized wave packet expands coherently across the lattice, generating a pronounced light-cone–like structure characteristic of ballistic transport in an extended phase. Upon increasing the interaction to $U=5$, the spreading is strongly suppressed and the wave packet remains largely confined near its initial site, indicating the onset of partial localization. For $U=7$, the dynamics is dominated by strong confinement, with the wave packet remaining localized throughout the evolution, signaling a fully localized regime. Interestingly, at larger interaction strength ($U=9$), the wave packet again exhibits appreciable spatial spreading, suggesting a re-emergence of extended behavior driven by interaction-induced restructuring of the effective single-particle states.
\begin{figure}[t]
    \centering
   \includegraphics[width=0.6\linewidth]{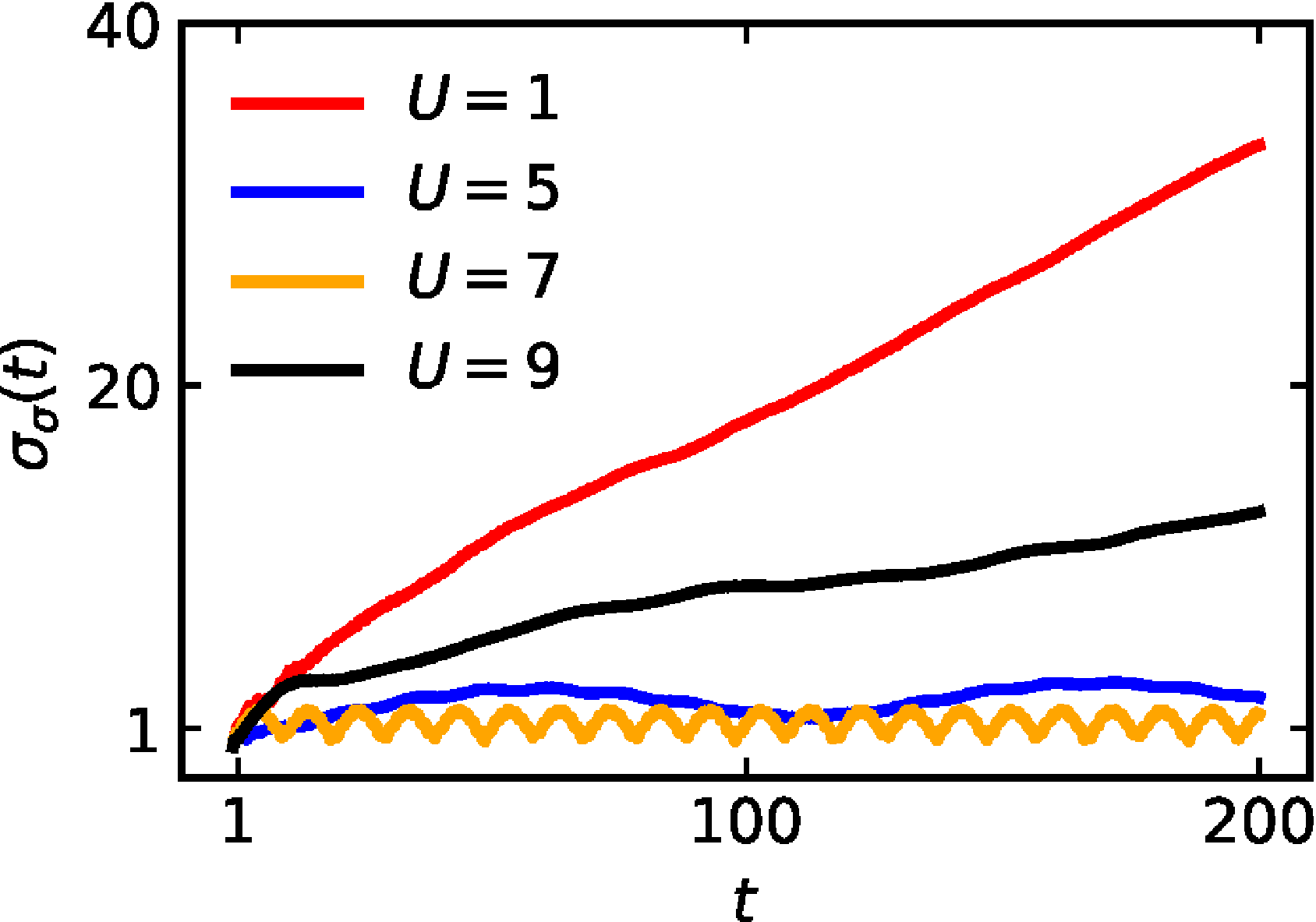}
    \caption{Time evolution of the root-mean-square displacement $\sigma(t)$ as a function of time for interaction strengths $U=1$, $5$, $7$, and $9$ at $\lambda=0.8$ and $h_z=0.8$.}
    \label{fig:dyna2}
\end{figure}
 \begin{figure}[t]
    \centering
   \includegraphics[width=1.0\linewidth]{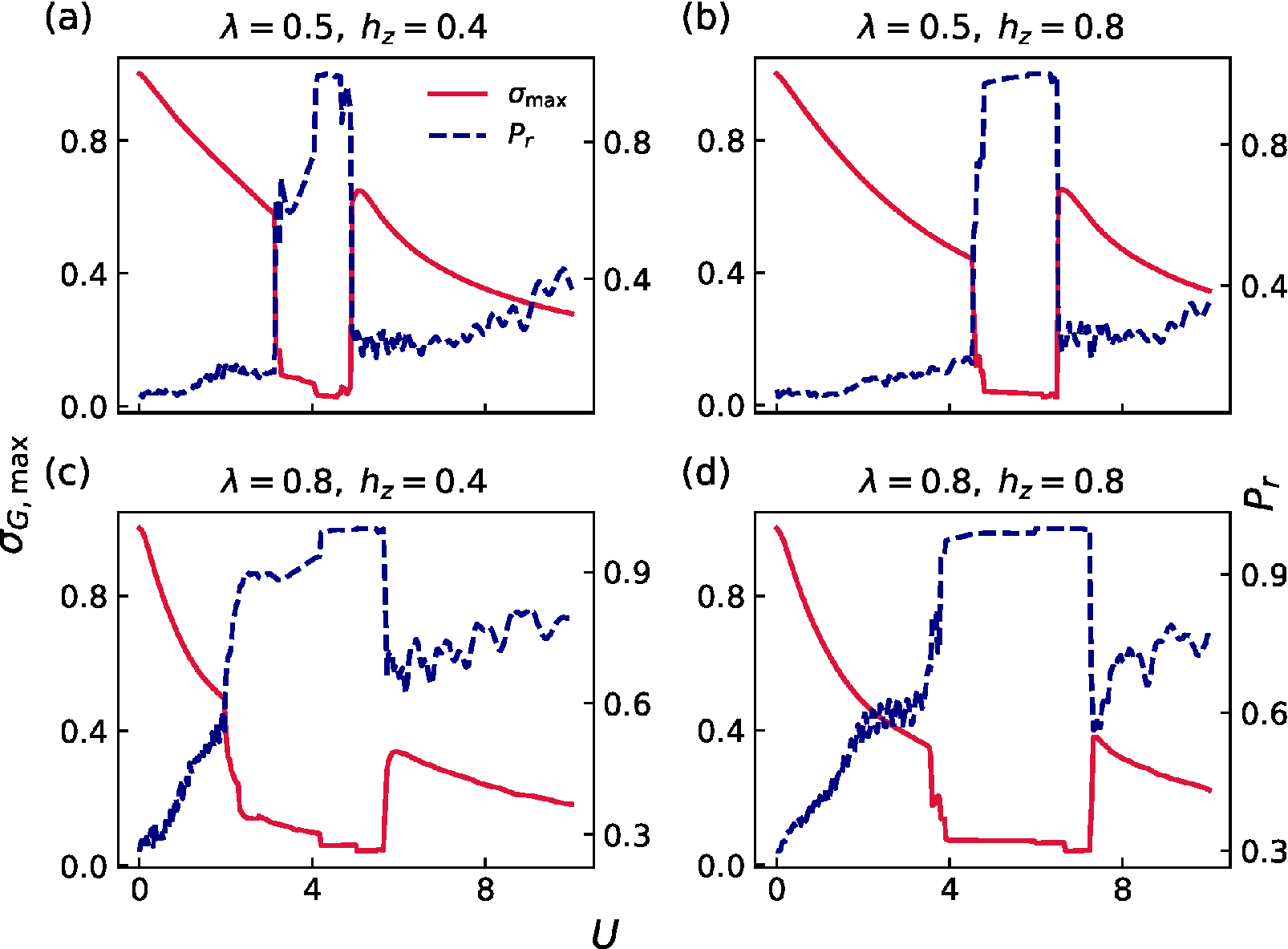}
    \caption{Maximum wave-packet spread $\sigma_{G,\max}$ (crimson solid line) and survival probability $P_r$ at $r=10$ (blue dashed line) plotted as functions of $U$ for four representative $\lambda$–$h_z$ parameter sets, as specified in Fig.~\ref{fig:uegnval}.}
    \label{fig:dyna3}
\end{figure}
To quantify the nonequilibrium dynamics of the interacting system, we analyze two complementary observables constructed from the spin-resolved time-dependent wave functions $\psi_{j,\sigma}(t)$. The first quantity is the root-mean-square (RMS) displacement~\cite{dyna3} of the wave packet, defined for each spin sector as
\begin{equation}
\sigma_{\sigma}(t)
= \left[
\sum_{j=1}^{L} (j-j_{0})^{2}
\left| \psi_{j,\sigma}(t) \right|^{2}
\right]^{1/2},
\end{equation}
where the initial excitation is placed at the central site $j_{0}=L/2$. Our numerical analysis shows that the temporal evolution of $\sigma_{\uparrow}(t)$ and $\sigma_{\downarrow}(t)$ follows nearly identical trends over the entire parameter range. We therefore characterize the wave-packet spreading using the spin-averaged displacement,
$\sigma(t) = \tfrac{1}{2}[\sigma_{\uparrow}(t)+\sigma_{\downarrow}(t)]$.

In the localized regime, the wave packet remains confined close to its initial position, causing the root-mean-square displacement $\sigma(t)$ to rapidly saturate at a small value. In contrast, both extended and intermediate regimes are characterized by appreciable spatial spreading, which leads to significantly larger saturation values of $\sigma(t)$. As shown in Fig.~\ref{fig:dyna2}, $\sigma(t)$ attains enhanced values for $U=1$ and $U=9$, signaling extended transport in these regimes. On the other hand, for $U=5$ and $U=7$ the saturation values are markedly reduced, with the smallest $\sigma(t)$ occurring at $U=7$, clearly indicating strong localization. These dynamical trends are fully consistent with the localization characteristics inferred from the static spectral and eigenstate analyses discussed earlier.

We denote the long-time limit by $\sigma_{t}$ and introduce a normalized measure $\sigma_{G,\max}=\sigma(t)/\sigma_{\max}$, where $\sigma_{\max}$ corresponds to the fully extended reference case (obtained for $U=0$). Consequently, $\sigma(t)/\sigma_{\max} \approx 1$ signals extended dynamics, $\sigma(t)/\sigma_{\max} \approx 0$ indicates strong localization, and intermediate finite values characterize partially extended behavior.

As a second diagnostic, we compute the spin-resolved survival probability~\cite{dyna4},
\begin{equation}
P_{r,\sigma}(t)
=
\sum_{j=\lceil L/2 \rceil - r}^{\lceil L/2 \rceil + r}
\left| \psi_{j,\sigma}(t) \right|^{2},
\end{equation}
which measures the likelihood that the wave packet remains within a finite window of width $2r+1$ centered around the initial site. Similar to the RMS displacement, the survival probabilities for spin-up and spin-down components exhibit almost indistinguishable temporal behavior, allowing us to define a spin-averaged quantity
$P_r(t)=\tfrac{1}{2}[P_{r,\uparrow}(t)+P_{r,\downarrow}(t)]$. In the long-time limit, $P_r(t)$ vanishes in the extended phase due to complete delocalization of the excitation, approaches unity in the localized phase where the wave packet remains trapped, and attains finite nonzero values in the intermediate regime. Together, the combined behavior of $\sigma(t)$ and $P_r(t)$ provides a robust and consistent characterization of the dynamical transport properties across different interaction-driven phases.

In Fig.~\ref{fig:dyna3}, we present the interaction dependence of the survival probability $P_r$ and the normalized saturation width $\sigma_{G,\max}$ for different parameter regimes, chosen consistently with the earlier figures. Both quantities exhibit trends that closely mirror those observed for the average IPR and average NPR in Fig.~\ref{fig:uainpr}. In particular, regimes characterized by enhanced localization (delocalization) correspond to larger (smaller) values of $P_r$ and reduced (enhanced) values of $\sigma_{G,\max}$. This close correspondence between dynamical and static diagnostics provides further confirmation that the interaction-driven localization behavior inferred from the eigenstate analysis is robust and persists under real-time evolution.


\section{Conclusion}

In summary, we have investigated the interplay of quasiperiodicity, electron-electron interactions, and a staggered spin-dependent Zeeman field in a spinful Hubbard ring within a self-consistent Hartree-Fock framework. Using complementary spectral, localization, multifractal, and real-space diagnostics, we find a pronounced nonmonotonic evolution of the electronic states with increasing Hubbard interaction. The system evolves from an extended regime to an intermediate localized/critical regime and subsequently undergoes re-entrant delocalization at stronger interactions. This evolution is consistently captured by the inverse participation ratio, normalized participation ratio, and fractal dimension, while the concomitant changes in charge-density fluctuations, local spin polarization, double occupancy, excitation gaps, and local entropy reveal the associated reconstruction of the self-consistent electronic structure. The intermediate regime broadens systematically with increasing quasiperiodic hopping modulation and staggered Zeeman-field strength, demonstrating their cooperative role in stabilizing interaction-driven criticality.

The spin-dependent field further leads to a marked spin selectivity of the interaction-induced reconstruction. In the intermediate regime, the self-consistent Hartree potentials generate distinct spectral and localization properties for the spin-up and spin-down sectors, as reflected in their spin-resolved energy spectra, IPRs, and multifractal characteristics. This asymmetry is accompanied by corresponding changes in the local charge and spin distributions and other self-consistent observables, establishing a direct connection between spin-dependent spectral reconstruction and the underlying real-space electronic structure. Thus, the combined action of correlations and the staggered Zeeman field provides a mechanism for controlling localization differently in the two spin channels.

The static picture is further corroborated by real-time wave-packet dynamics, which exhibit ballistic propagation, spatial confinement, and re-entrant transport across the successive regimes identified from the eigenstate analysis. These results demonstrate that quasiperiodicity, correlations, and spin-dependent fields jointly determine the localization and transport properties of the system, with electronic interactions playing a dual role in promoting localization at intermediate coupling and restoring delocalization at stronger coupling. Our findings provide a unified picture of interaction-driven and spin-selective localization in correlated quasiperiodic systems and motivate further studies beyond the Hartree-Fock level, including quantum fluctuations, longer-range interactions, dissipation, and higher-dimensional quasiperiodic architectures.

\bibliography{ref}

\end{document}